\documentclass[twocolumn,showpacs,aps,prd,superscriptaddress]{revtex4-1}
\usepackage{microtype}  
\usepackage{hyperref}

\usepackage{amsmath}
\usepackage{graphicx}
\usepackage[usenames]{color}
\usepackage{xspace}
\usepackage{float}

\usepackage[caption=false]{subfig}

\newcommand{\beq}{\begin{equation}}
\newcommand{\eeq}{\end{equation}}

\newcommand{\beqn}{\begin{eqnarray}}
\newcommand{\eeqn}{\end{eqnarray}}

\newcommand{\nrpy}{{\texttt{NRPy}}\xspace}
\newcommand{\TwoPunctures}{{\texttt{TwoPunctures}}\xspace}
\newcommand{\Baikal}{{\texttt{Baikal}}\xspace}
\newcommand{\BaikalVacuum}{{\texttt{BaikalVacuum}}\xspace}
\newcommand{\et}{\texttt{Einstein Toolkit}\xspace}
\newcommand{\Carpet}{\texttt{Carpet}\xspace}
\newcommand{\Llama}{\texttt{Llama}\xspace}

\newcommand{\bhah}{{\texttt{BlackHoles@Home}}\xspace}
\newcommand{\qlm}{{\texttt{QuasiLocalMeasures}}\xspace}
\newcommand{\kranc}{{\texttt{Kranc}}\xspace}
\newcommand{\weylscal}{{\texttt{WeylScal4}}\xspace}
\newcommand{\ahfd}{{\texttt{AHFinderDirect}}\xspace}
\newcommand{\control}{{Control}\xspace}
\newcommand{\cahd}{{CA${\mathcal H}$D}\xspace}
\newcommand{\cako}{{CAKO}\xspace}
\newcommand{\ssl}{{SSL}\xspace}
\newcommand{\allthree}{{All3}\xspace}

\renewcommand{\eqref}[1]{Eq.\,(\ref{#1})}

\begin{document}
\date{\today}
\title{Improved Moving-Puncture Techniques for Compact Binary Simulations}

\author{Zachariah B.~Etienne}
\email{zetienne@uidaho.edu}
\affiliation{Department of Physics, University of Idaho, Moscow, ID 83843, USA}
\affiliation{Department of Physics and Astronomy, West Virginia University, Morgantown, WV 26506, USA}
\affiliation{Center for Gravitational Waves and Cosmology, West Virginia University, Chestnut Ridge Research Building, Morgantown, WV 26505, USA}

\begin{abstract}
  To fully unlock the scientific potential of upcoming gravitational wave (GW) interferometers, numerical relativity (NR) simulation accuracy will need to be greatly enhanced. We present three infrastructure-agnostic improvements to the moving-puncture approach for binary black hole (BBH) simulations, aimed at greatly reducing constraint violation and improving GW predictions. Although these improvements were developed within the highly efficient NR code \bhah, we demonstrate their effectiveness in the widely-adopted \et/\Carpet AMR framework. Our improvements include a modified Kreiss-Oliger dissipation prescription, a Hamiltonian-constraint-damping adjustment to the BSSN equations, and an extra term to the 1+log lapse evolution equation that slows the development of the sharp lapse feature, which dominates numerical errors in BBH simulations. With minimal increase in computational cost, these improvements greatly reduce GW noise, enabling the extraction of high-order GW modes previously obscured by numerical noise. They also improve convergence properties near and inside the convergent regime, reduce Hamiltonian (momentum) constraint violations in the strong-field region by roughly two (three) orders of magnitude, and in the GW-extraction zone by five (two) orders of magnitude. To promote community adoption, we have open-sourced the improved \et thorn \BaikalVacuum used in this work. Although our focus is on BBH evolutions and the BSSN formulation, these improvements may also benefit compact binary simulations involving matter and other formulations, a focus for future investigations.
\end{abstract}

\maketitle

\section{Introduction}
\label{sec:intro}

Numerical relativity (NR) simulations of compact binary mergers play a crucial role in gravitational wave (GW) data analysis, enabling reliable estimation of physical parameters from the late-inspiral, merger, and post-merger phases of GW signals. To date, detector noise in GW observations has generally surpassed the numerical errors intrinsic to NR simulations and GW approximants derived from them, so GW parameter estimation errors have generally been dominated by noise in the GW signal. However, improved sensitivities offered by current and forthcoming GW detectors threaten to change this dynamic. Third-generation (3G) detectors will necessitate a roughly tenfold improvement in NR simulation accuracy~\cite{Purrer:2019jcp,Ferguson:2020xnm}. Therefore, enhancing the accuracy of NR codes is vital for unlocking the full scientific potential of GW astronomy moving forward.

Improved accuracy can be achieved through various means. One widely adopted approach involves building next-generation NR codes upon more scalable infrastructures. Increased scalability enables simulations to be performed at higher resolutions and with greater physical realism. While running simulations at higher resolutions generally reduces numerical errors, it does not directly address one of the primary sources of error in existing NR codes: noise associated with the selection of numerical grids.

Cartesian adaptive mesh refinement (AMR) grids, the most popular gridding approach in NR, span the many decades in physical scales intrinsic to compact binaries with nested Cartesian grid patches. The highest resolution patches are placed where the spacetime fields are sharpest: at the compact objects themselves. These patches are embedded in increasingly coarser patches, extending to the outer boundary. Cartesian AMR grids are both numerically convenient and highly scalable with modern AMR algorithms.

When it comes to solving Einstein's equations in NR, the moving-puncture approach~\cite{Campanelli:2005dd,Baker:2005vv} is the most popular method. This approach generally combines the BSSN~\cite{Baumgarte:1998te,Shibata:1995we}, CCZ4~\cite{Alic:2011gg} or Z4c~\cite{Bernuzzi:2009ex} 3+1 decompositions of Einstein's equations with the moving-puncture gauge conditions~\cite{Alcubierre:2002kk,Campanelli:2005dd,Baker:2005vv}, enabling the evolution of black holes (BHs) that traverse the numerical grid without the need to excise their interiors. However, mixing the moving-puncture technique with Cartesian AMR grids results in significant numerical noise that negatively impacts the accuracy of compact binary simulations.

\begin{figure*}
  \includegraphics[angle=0,width=0.45\linewidth]{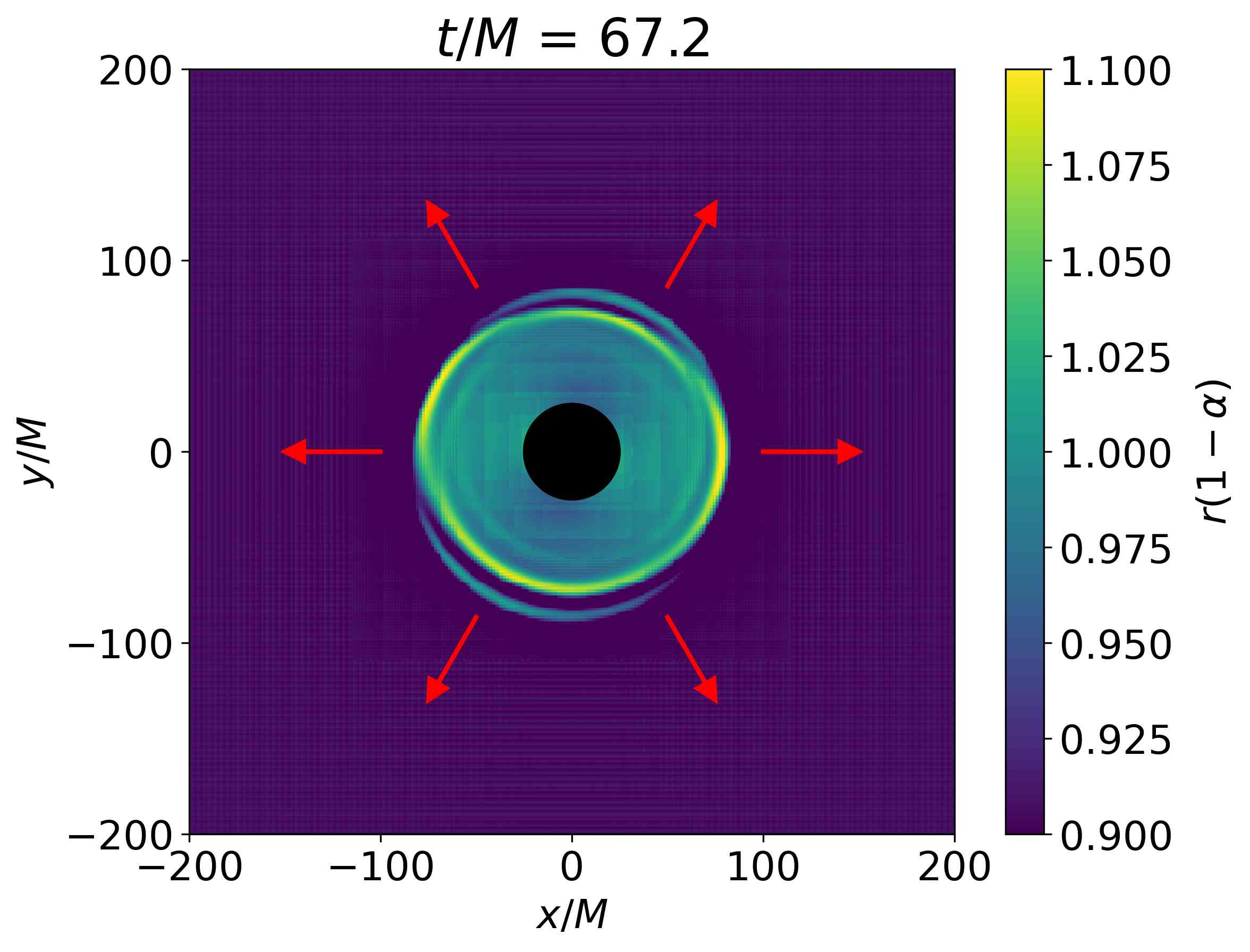} \hspace{0.8cm}
  \includegraphics[angle=0,width=0.45\linewidth]{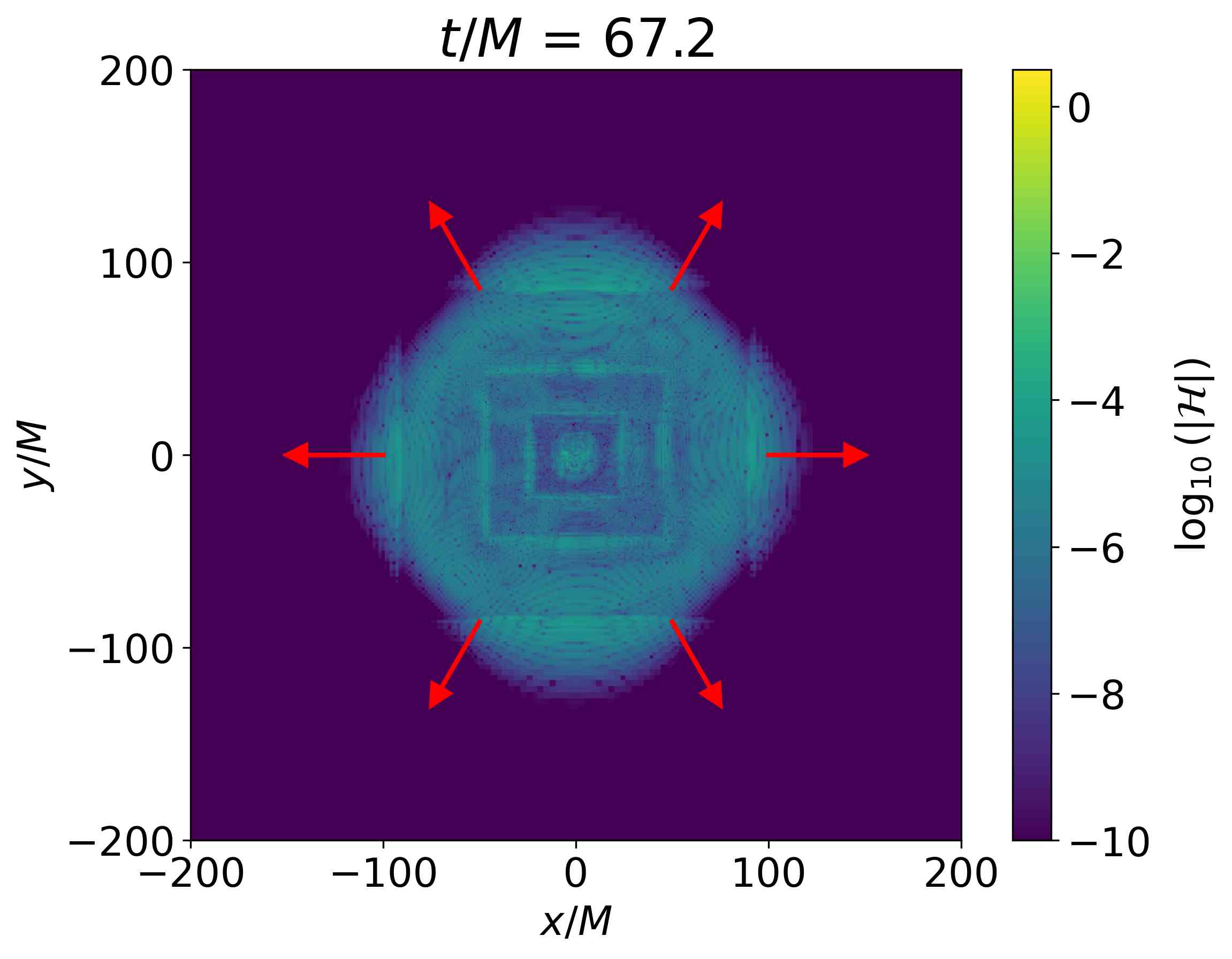} \\
  \includegraphics[angle=0,width=0.45\linewidth]{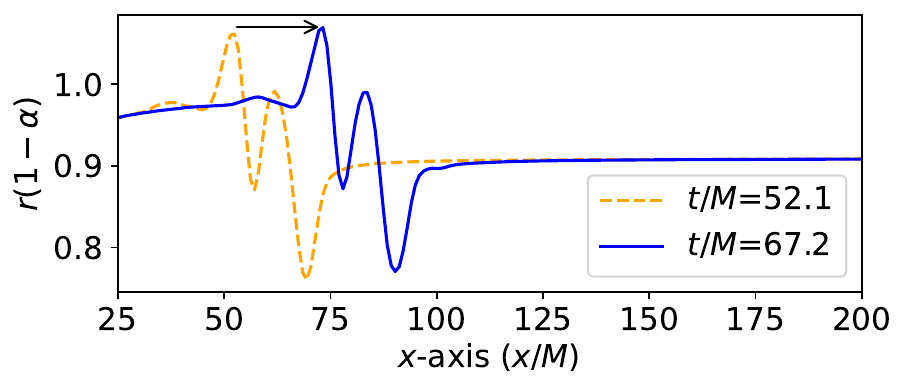} \hspace{0.8cm}
  \includegraphics[angle=0,width=0.45\linewidth]{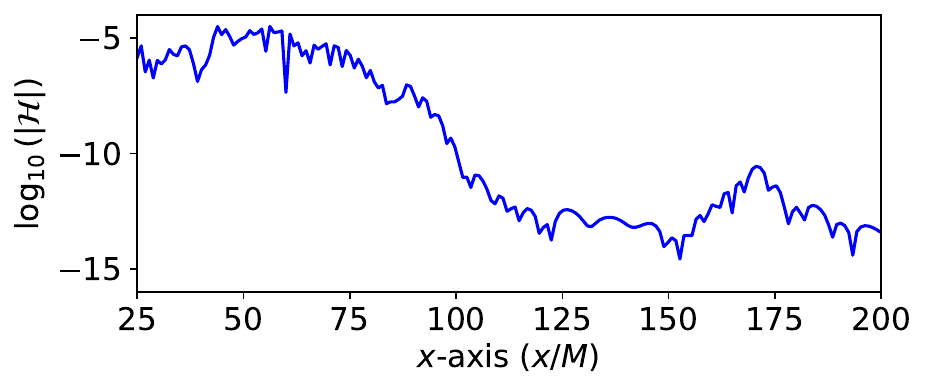}
  \caption{\textbf{Development of sharp moving-puncture lapse feature and its aftermath.} \textbf{Top left}: $r(1-\alpha)$ colormap on the orbital plane, for $r>25M$; data for $r\leq 25M$ were excised (represented by a black dot). \textbf{Bottom left}: $r(1-\alpha)$ data along the $x$-axis, for $x>25M$. \textbf{Top right}: $\mathcal{H}$ constraint violation, plotted as $\log_{10}|\mathcal{H}|$ on the orbital plane. The dashed orange curve corresponds to the data in the top-left plot, and the blue solid curve corresponds to the data at $t=15.1M$ later. The black arrow denotes the propagation distance and direction of a wave propagating outward at $\sqrt{2}c$, from $t/M=52.1$ to $t/M=67.2$. \textbf{Bottom right}: $\log_{10}|\mathcal{H}|$ along the $x$-axis at $t=67.2M$, for $x>25M$.}
  \label{fig:sharplapse}
\end{figure*}

As shown in Fig.~\ref{fig:sharplapse}, during the initial phase of a moving-puncture binary black hole (BBH) evolution, the lapse rapidly settles to new values in the strong-field region, releasing a sharp, superluminal gauge wave packet that radiates outward from the BHs (top panel) at a speed of $\sqrt{2}c$---the characteristic speed of the 1+log lapse~\cite{Etienne:2014tia} and the only superluminal characteristic speed~\cite{vanMeter:2006vi}. Given our standard choice of initial (``precollapsed'') lapse, in the weak field the lapse tends to a value of $1-\alpha \propto r^{-1}$, both before and after the wave packet passes. Thus away from the wave packet, the $r(1-\alpha)$ plotted in Fig.~\ref{fig:sharplapse} tends to constant values in the weak field.

As the sharp lapse wave packet travels outward from the strong-field region, it crosses multiple AMR refinement boundaries. At each boundary, the numerical resolution drops discontinuously by a factor of two~\footnote{In virtually all AMR implementations, and even for those that offer refinement factors other than two, two is generally used without exception.}. This crossing partially reflects the wave and generates substantial numerical noise in its wake.

Although this is a gauge wave, errors in its evolution directly contaminate the evolution of physical quantities, particularly the extrinsic curvature, which depends on second-order derivatives of the lapse. As Einstein's equations are highly nonlinear, errors in the lapse's evolution contaminate all physical quantities, including the Einstein constraints. The right panel of Fig.~\ref{fig:sharplapse} illustrates the roughly seven-decade jump in the Hamiltonian-constraint violation $\mathcal{H}$ from its initial values following the passage of this wave. Elevated levels of $\mathcal{H}$ violation persist near refinement boundaries, evident as square shapes in the colormap. 

Meanwhile, in the strong-field region, $\mathcal{H}$ violation acts as an effective energy density, directly influencing the dynamics of the compact binary and, thus, one of the core outputs of the NR simulation: the GW prediction. 
Regarding GW predictions, noise appears as spurious high-frequency oscillations in the outgoing-radiation Weyl scalar $\psi_4$. $\psi_4$ is decomposed into spin-weight $s=-2$ spherical harmonics and integrated twice in time to compute the GW strain. While the noise is evident in the ($\ell=m=2$) dominant mode of $\psi_4$, it completely obscures modes with $\ell > 4$. Such high-order modes may be detectable by next-generation GW observatories, even in near-symmetric binaries~\cite{LISA:2017pwj}. Thus minimizing this noise is crucial for precise BBH waveform predictions for upcoming GW detectors.

We emphasize that this discussion has been limited to Cartesian AMR grids not set to refine at the sharp gauge features, as this approach is widely used in the field. Improved refinement algorithms or grid structures may result in significantly less constraint violation and GW noise than what is reported here.

This paper introduces improvements to the moving-puncture approach, aimed at curtailing the development of sharp features in the lapse and mitigating their impact. Our first two improvements focus on decreasing the strength of low-pass filters near punctures---where sharp fields are desirable---and increasing the strength far away.

The first improvement multiplies the strength parameter of Kreiss-Oliger (KO) dissipation by $e^{-4\phi}$, where $\phi$ is the standard BSSN conformal factor, and increases its strength from typical values by 2 to 5 times.

The second improvement modifies a powerful but not widely used low-pass parabolic filter, which adds a term proportional to $\mathcal{H}$ to the right-hand side of the $\phi$ evolution equation. In the same vein as \cite{Assumpcao:2021fhq}, we set the strength of the filter based on the maximum value allowed by the CFL condition, according to the local numerical grid spacing. This enables us to increase its strength by more than an order of magnitude on coarser numerical grids in the region far from the punctures.

Our third improvement adds a term to the 1+log lapse evolution equation that slows the evolution of $\alpha$ in the area immediately outside the BH horizon. This arrests the formation of the sharp lapse wave packet, both reducing its magnitude and greatly smoothing it.

Modifying the moving-puncture approach in Cartesian-AMR BBH evolutions to mitigate the impact of the sharp lapse feature is not a new idea; \cite{Etienne:2014tia} introduced several techniques to this end, achieving about a 40\% reduction in GW noise (in $\psi_4$) and about 1.3 orders of magnitude reduction in $|\mathcal{H}|$ violations outside the horizons. While these improvements were effective, their recipes depend on careful tuning to the specific grid structure.

While we adopt the BSSN formalism (like~\cite{Etienne:2014tia}) and employ a constraint-damping term, the moving-puncture approach is also widely used within the CCZ4 and Z4c formalisms, which contain constraint-damping terms. \cite{Hilditch:2012fp} performed evolutions of puncture BBHs in the Z4c formalism, reporting 1--2 order of magnitude reductions in Hamiltonian constraint violation outside the punctures, and GW phase and amplitude errors up to a factor of two smaller as compared to the standard BSSN moving-puncture approach.

By contrast, the three improvements presented in this work, when combined, reduce Hamiltonian (momentum) constraint violations in the strong-field region by about two (three) orders of magnitude and in the GW-extraction zone by five (two) orders of magnitude. Further, constraint violation convergence is generally improved, implying that these gaps widen with increased resolution. Coordinate eccentricity is slightly decreased, and numerical noise in the dominant $\ell=m=2$ mode of $\psi_4$ is reduced by an average factor of 4.3 across resolutions. Fidelity of higher-order modes is greatly improved, enabling the $\ell=m=6$ mode, previously obscured by noise, to become visible.

Although our focus is primarily on the impact of these improvements on BBH evolutions with the BSSN formalism, we expect that they will also be beneficial in other compact binary evolutions and formalisms that use the moving-puncture technique. Determining whether our improved techniques significantly improve the fidelity of binary neutron star (BNS) simulations will be a focus of an upcoming paper. In addition, formulations like Z4c have proven quite effective in reducing constraint violations in BNS simulations~\cite{Hilditch:2012fp}. Comparing BNS evolutions performed with our improved BSSN moving-puncture approach to Z4 techniques widely used across the community will also be a focus of future work.

The rest of the paper is structured as follows. Section~\ref{sec:basiceqs} introduces the basic equations, including the BSSN evolution, moving-puncture gauge, and constraint equations. Building on this, Sec.~\ref{sec:improvedtechniques} presents prescriptions for all three of our improved techniques. Next, Sec.~\ref{sec:alg_approach} describes our algorithmic approach for simulating the BBH mergers presented in this work. Results from these simulations, discussed in Sec.~\ref{sec:results}, demonstrate the impacts of our improved techniques. Finally, we conclude and describe avenues for future work in Sec.~\ref{sec:conclusion}.

\section{Basic Equations}
\label{sec:basiceqs}

We adopt $G=c=1$ units, and set our mass scale $M=1$ to be the sum of the individual puncture masses reported by our initial data solver \TwoPunctures. In these units the total ADM mass is 0.989946. Further we adopt standard Einstein summation rules on repeated indices, and the convention that Greek and Latin indices represent 4- and 3-dimensional quantities, respectively.

\subsection{BSSN Evolution Equations}
\label{subsec:BSSNevolution}

In this paper, we investigate the numerical evolutions of BBHs using the covariant ``Lagrangian'' BSSN formulation in vacuum, as presented in~\cite{BrownCovariantBSSN}. This formulation has been adapted for spherical coordinate reference metrics by~\cite{Baumgarte:2012xy} and for more general reference metrics by~\cite{Ruchlin:2017com}.

Starting with the ADM line element,
\beq
ds^2 = -\alpha^2 dt^2 + \gamma_{ij}\left(dx^i + \beta^i dt\right)\left(dx^j + \beta^j dt\right),
\eeq
$\gamma_{ij}$, $\beta^i$, and $\alpha$ denote the spatial 3-metric, shift vector, and lapse, respectively.

We then define the conformal metric $\bar{\gamma}_{ij}$ as
\beq
\bar{\gamma}_{ij} = \left(\frac{\bar{\gamma}}{\gamma}\right)^{\frac{1}{3}} \gamma_{ij} = e^{-4\phi} \gamma_{ij},
\eeq
where $\phi$ represents the conformal factor. Instead of evolving $\phi$, we follow previous works (e.g.,~\cite{Tichy:2007hk,Marronetti:2007wz}) and evolve $W=e^{-2\phi}$, as $W$ exhibits greater smoothness near puncture BHs, resulting in reduced truncation errors (also see discussion in~\cite{Campanelli:2005dd}). To this end, derivatives of the conformal factor $\phi$ appearing in the BSSN evolution equations are calculated with respect to the evolved variable $W$ via $\partial_\mu W = -2 e^{-2\phi} \partial_\mu \phi \Rightarrow \partial_\mu \phi = -\partial_\mu W / (2W)$, and so forth for second derivatives.

We also introduce a reference metric $\hat{\gamma}_{ij}$, defined by
\beq
\bar{\gamma}_{ij} = \hat{\gamma}_{ij} + \varepsilon_{ij}.
\eeq
In this work, the reference metric corresponds to the Cartesian metric $\hat{\gamma}_{ij} = \delta_{ij}$, which remains constant in both space and time. Thus, all derivatives of the conformal metric $\bar{\gamma}_{ij}$ are evaluated directly with respect to $\varepsilon_{ij}$, since $\partial_\mu \bar{\gamma}_{ij} = \partial_\mu \varepsilon_{ij}$. Although analytically equivalent, computing finite-difference derivatives of diagonal metric components with respect to $\varepsilon_{ij}$ instead of $\bar{\gamma}_{ij} = 1 + \varepsilon_{ij}$ results in reduced roundoff error in the weak-field region.

Building on these definitions, we adopt the following time-evolution equations, replacing all occurrences of $\phi$ and its derivatives with equivalent expressions in terms of $W$:
\beqn
\partial_{t} \varepsilon_{i j} &=& \mathcal{L}_\beta \varepsilon_{i j} + \frac{2}{3} \bar{\gamma}_{i j} \left (\alpha \bar{A}_{k}^{k} - \bar{D}_{k} \beta^{k}\right ) - 2 \alpha \bar{A}_{i j} \\
\partial_{t} \bar{A}_{i j} &=& \mathcal{L}_\beta \bar{A}_{i j} -\frac{2}{3} \bar{A}_{i j} \bar{D}_{k} \beta^{k} - 2 \alpha \bar{A}_{i k} {\bar{A}^{k}}_{j} + \alpha \bar{A}_{i j} K \nonumber \\
&& + e^{-4 \phi} \left \{-2 \alpha \bar{D}_{i} \bar{D}_{j} \phi + 4 \alpha \bar{D}_{i} \phi \bar{D}_{j} \phi \right . \nonumber \\
&& \left . + 4 \bar{D}_{(i} \alpha \bar{D}_{j)} \phi - \bar{D}_{i} \bar{D}_{j} \alpha + \alpha \bar{R}_{i j} \right \}^{\text{TF}} \\
\label{eq:phievol}
\partial_t \phi &=& \mathcal{L}_\beta \phi + \frac{1}{6} \left (\bar{D}_{k} \beta^{k} - \alpha K \right )  \\
%
\label{eq:Kevol}
\partial_{t} K &=& \mathcal{L}_\beta K + \frac{1}{3} \alpha K^{2} + \alpha \bar{A}_{i j} \bar{A}^{i j} \nonumber \\
&& - e^{-4 \phi} \left (\bar{D}_{i} \bar{D}^{i} \alpha + 2 \bar{D}^{i} \alpha \bar{D}_{i} \phi \right ) \\
\partial_{t} \bar{\Lambda}^{i} &=& \mathcal{L}_\beta \bar{\Lambda}^{i} + \bar{\gamma}^{j k} \hat{D}_{j} \hat{D}_{k} \beta^{i} + \frac{2}{3} \Delta^{i} \bar{D}_{j} \beta^{j} \nonumber \\
&& + \frac{1}{3} \bar{D}^{i} \bar{D}_{j} \beta^{j}  - 2 \bar{A}^{i j} \left (\partial_{j} \alpha - 6 \partial_{j} \phi \right ) \nonumber \\
&&+ 2 \bar{A}^{j k} \Delta_{j k}^{i} -\frac{4}{3} \alpha \bar{\gamma}^{i j} \partial_{j} K
\eeqn
where the $\text{TF}$ superscript indicates the trace-free part, and $\mathcal{L}_\beta$ denotes the Lie derivative along the shift vector $\beta^i$. The covariant derivatives with respect to the reference metric $\hat{\gamma}_{ij}$ and the barred spatial 3-metric $\bar{\gamma}_{ij}$ are represented by $\hat{D}_j$ and $\bar{D}_j$, respectively. The tensor $\Delta^i_{jk}$ is formed from the difference between the barred and hatted Christoffel symbols:
$$\Delta^i_{jk} = \bar{\Gamma}^i_{jk} - \hat{\Gamma}^i_{jk},$$
where $\Delta^i$ is defined $\Delta^i \equiv \bar{\gamma}^{jk} \Delta^i_{jk}$. The conformal, trace-free extrinsic curvature, $\bar{A}_{ij}$, is given by:
$$\bar{A}_{ij} = e^{-4\phi} \left(K_{ij} - \frac{1}{3}\gamma_{ij} K\right),$$
where $K$ represents the trace of the extrinsic curvature $K_{ij}$.

\subsection{Moving-Puncture Gauge Evolution Equations}

To close the system of time-evolution equations, we choose the moving-puncture gauge conditions~\cite{Alcubierre:2002kk,Campanelli:2005dd,Baker:2005vv}, which include the 1+log slicing condition for the lapse and the $\Gamma$-driving (here, $\Lambda$ takes the place of $\Gamma$) condition for the shift:
\beqn
\label{eq:onepluslog}
\partial_t \alpha &=& \beta^j \partial_j \alpha -2\alpha K \\
\partial_t \beta^i &=& \beta^j \partial_j \beta^i + B^i \\
\partial_t B^i &=& \beta^j \partial_j B^i + \frac{3}{4} \left(\partial_t \bar{\Lambda}^i + \beta^j \partial_j\bar{\Lambda}^i\right) - \eta \beta^i,
\eeqn
where the $\Gamma$-driving shift damping parameter is set in this paper to $\eta=1/M$.

Variations in moving-puncture gauge conditions, such as a first-order implementation of the shift condition, do exist. Although we did not explore these alternatives in our study, we direct the reader to van Meter \textit{et al}.~\cite{vanMeter:2006vi} for a comprehensive analysis of different implementations. Our choice of gauge conditions was primarily motivated by their finding that incorporating advection terms into the lapse and shift conditions leads to well-behaved characteristic speeds.


\subsection{BSSN Constraint Equations}
\label{subsec:BSSNconstraint}

Analytically (as opposed to numerically), the BSSN Hamiltonian constraint $\mathcal{H}=0$, so any expression for $\mathcal{H}$ can be multiplied by an arbitrary prefactor and still remain perfectly valid. As such, two expressions for the BSSN Hamiltonian constraint $\mathcal{H}$ are commonly found in the literature, differing only in their prefactors.

In this work, we adopt $\mathcal{H}$ with a negative prefactor in front of $\bar{D}^2 \phi$, its principal part. To differentiate our definition of $\mathcal{H}$ from other choices, we will refer to it as $\mathcal{H}_{-}$ when necessary:
\beqn
\mathcal{H}_{-} &=& \frac{2}{3} K^2 - \bar{A}_{ij} \bar{A}^{ij} +
e^{-4\phi} \left(\bar{R} - 8 \bar{D}^i \phi \bar{D}_i \phi - 8 \bar{D}^2 \phi\right) \nonumber \\
&=& \mathcal{H} = 0.
\eeqn

The second convention (e.g.,~\cite{Duez:2002bn}) multiplies both sides by $-e^{5\phi}/8$. This results in a sign flip on $\mathcal{H}$, and thus the prefactor on $\bar{D}^2 \phi$ becomes positive. We therefore refer to this form as $\mathcal{H}_{+}$:
\beq
\mathcal{H}_{+} = -\frac{e^{5\phi}}{8} \mathcal{H}_{-}.
\eeq

Although the two expressions $\mathcal{H}_{+}=\mathcal{H}_{-}=0$ are equivalent, it is important to distinguish between them. As we will soon see, the specific implementation of Hamiltonian-constraint damping (Sec.~\ref{subsec:Hconstdamping}) will depend on the choice of this prefactor.

Regarding the physical implications of constraint violation, if we write $\mathcal{H}_{-}$ in its general form with the matter source term included:
\beqn
\mathcal{H}^m_{-} &=& \mathcal{H}_{-} - 16\pi \rho = 0,
\eeqn
then we may conclude that, in vacuum evolutions, positive Hamiltonian-constraint violations $\mathcal{H}_{-}$ behave as a positive \textit{effective} energy density $\rho_{\rm eff}$, and negative violations as a negative $\rho_{\rm eff}$:
\beq
\mathcal{H}_{-} = 16\pi \rho_{\rm eff}.
\label{eq:effectivedensity}
\eeq

As for the momentum constraint, we adopt the following form:
\beqn
\mathcal{M}^{i} &=& e^{-4 \phi} \left (\hat{D}_{j} \bar{A}^{i j} + 2 \bar{A}^{k (i} \Delta^{j)}_{j k} + 6 \bar{A}^{i j} \partial_{j} \phi - \frac{2}{3} \bar{\gamma}^{i j} \partial_{j} K \right ) \nonumber \\
&=& 0 \; ,
\eeqn
and define the scalar $\mathcal{M}^2 = \bar{\gamma}_{ij} \mathcal{M}^i \mathcal{M}^j$, which we use for diagnostics in this work.

As Hamiltonian- and momentum-constraint violation are defined at each of the $\gtrsim 10^7$ points in our computational domain, for convenience we will often compute them as a volume-averaged $L_2$ norm:
\beqn
||\mathcal{H}|| &=& \sqrt{\frac{\int_{\mathcal{V}} \mathcal{H}^2 d^3x}{\int_{\mathcal{V}} d^3x}} \\
\sqrt{||\mathcal{M}^2||} &=& \sqrt[4]{\frac{\int_{\mathcal{V}} (\bar{\gamma}_{ij} \mathcal{M}^i \mathcal{M}^j)^2 d^3x}{\int_{\mathcal{V}} d^3x}},
\eeqn
where the volume $\mathcal{V}$ will generally exclude small spheres around the BHs and be tailored to either a strong-field or weak-field region.

\section{Improved Techniques}
\label{sec:improvedtechniques}

We present three techniques that lower numerical errors in moving-puncture simulations, with minimal additional computational cost. These include curvature-adjusted Kreiss-Oliger dissipation (\cako), coarse-grid-adjusted Hamiltonian-constraint damping (\cahd), and the slow-start lapse (\ssl) technique.

These improvements are meant to be both easy to implement and agnostic to numerical grid structures. To this end, while we demonstrate their effectiveness in the widely used \et~\cite{EinsteinToolkit,EinsteinToolkit:web} \Carpet~\cite{Carpet,CarpetCode:web} AMR infrastructure, the new techniques were originally devised and tuned during the development of a new, super-efficient NR code called \bhah.

\begin{figure}[ht!]
  \begin{center}
    \vspace{-0.cm}
    \includegraphics*[angle=0,width=0.48\textwidth]{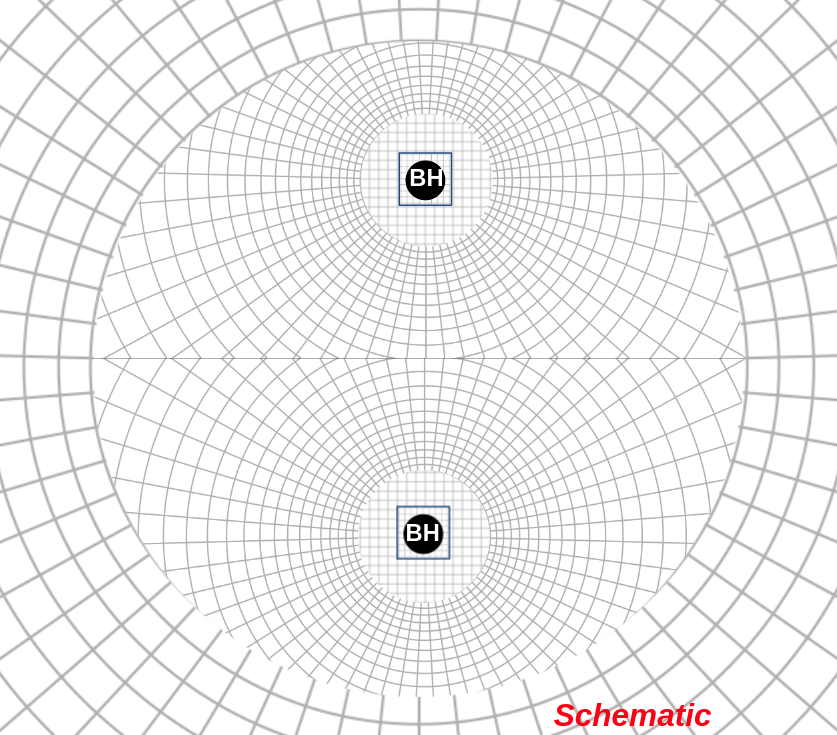}
  \end{center}
  \vspace{-0.4cm}
  \captionsetup{width=0.5\textwidth}
  \caption{Schematic of BBH inspiral numerical meshes in \bhah, the code used to develop and tune the improvements presented here.}
  \label{fig:bhahgrids}
\end{figure}

Illustrated in Fig.~\ref{fig:bhahgrids}, \bhah adopts a unique multi-patch grid structure that combines grids in spherical-like and Cartesian-like coordinates. Einstein's equations are solved in the corresponding spherical-like and Cartesian-like bases, leveraging recent developments in covariant BSSN~\cite{BrownCovariantBSSN}, BSSN with a reference metric~\cite{Baumgarte:2012xy,Ruchlin:2017com}, and NR code generation with \nrpy~\cite{Ruchlin:2017com,nrpy_web}.

With these new grid structures, \bhah performs full, 3+1 NR BBH evolutions using about 3GB of RAM or less, enabling its use on consumer-grade desktop computers. As such, it aims to generate the largest-ever full-NR BBH GW catalog through an upcoming volunteer-computing project of the same name.

Owing to \bhah's high efficiency, the improvements presented here were tuned over approximately 10 simulations of a single BBH scenario, and the tuned values were proven similarly effective across more than 100 different BBH scenarios simulated with \bhah. These scenarios spanned mass ratios from $q=1$ to $q=6$ and included arbitrarily directed spin vectors with dimensionless magnitudes up to 0.5.

Building on the success of these \bhah-tuned improvements, this work focuses on evaluating their performance on Cartesian AMR grids. To this end, we implement the improvements within our \BaikalVacuum \et thorn, and release the improved version of \BaikalVacuum used in this paper open source~\cite{BaikalVacuum2024}. This updated version of \BaikalVacuum and its sister (non-vacuum BSSN) code \Baikal will be incorporated into a future \et release.

We next describe each improvement in detail.

\subsection{Curvature-Adjusted Kreiss-Oliger dissipation (\cako)}
\label{subsec:improvedko}


Kreiss-Oliger (KO) dissipation~\cite{Kreiss73}, a widely adopted low-pass filter in numerical relativity (NR), reduces numerical noise. In Cartesian coordinates, the addition of the $n$th-order KO dissipation operator (where $n$ is odd) to the right-hand side of each evolution equation is formulated as:
\beq
D_{\mathrm{KO}}^n \vec{f} = \epsilon_{\mathrm{KO}} \frac{(-1)^{(n+3)/2}}{2^{n+1}} (\Delta x^i)^n \partial^{(n+1)}_i \vec{f},
\eeq
where $\vec{f}$ denotes the vector of evolved functions on a numerical grid, $\epsilon_{\mathrm{KO}}$ is the dimensionless KO strength factor, and $\partial^{(n+1)}_i$ represents the $(n+1)$st partial derivative in the $x^i$ direction. This derivative is generally implemented by a centered finite-difference operator with the smallest possible stencil. For example, first-order KO dissipation includes the second derivative, which is represented by the standard 3-point stencil: $\partial_x^2 f = (f_{i-1} - 2 f_i + f_{i+1})/(\Delta x)^2$. The $(\Delta x^i)^n$ prefactor ensures that the operator converges to zero in the continuum limit; KO dissipation is purely a numerical filter acting to damp high-frequency oscillations.

When solving a PDE with a dominant finite-difference truncation error of $j$th order ($j$ even), the $n = j + 1$-order KO dissipation operator is most often chosen, with the KO operator stencil comprising $j + 3$ grid points. In this study, we use $j = 8$, and apply 9th-order KO dissipation to the central point on an 11-point stencil.

Larger KO strength factors $\epsilon_{\mathrm{KO}}$ lead to stronger filtering; however, $\epsilon_{\mathrm{KO}}$ is CFL-limited to values of approximately one or less when typical CFL factors and numerical integration schemes are chosen for evolving Einstein's equations in the moving-puncture formalism~\cite{Bozzola:2021elc}. This said, so long as the CFL-limited value is not exceeded, $\epsilon_{\mathrm{KO}}$ need not be the same for all evolved grid functions.

In~\cite{Liu2009}, we found that setting $\epsilon_{\mathrm{KO,gauge}}=0.9$ for the gauge quantities and $\epsilon_{\mathrm{KO,other}}=0.3$ for the other spacetime quantities enhances stability when evolving boosted and unboosted near-extremal-spin ($a/M=0.99$) BHs. Several years later, in~\cite{Etienne:2014tia}, we refined this approach in the context of a mildly spinning BBH evolution. Building on the idea that the sharp fields inside and around puncture BHs are natural and should not be smoothed, we made $\epsilon_{\mathrm{KO,gauge}}$ a function of time and the distance from the center-of-mass $r$, such that it increased gradually over time from 0.3 to 0.98 in the weak-field region. Additionally, we adjusted $\epsilon_{\mathrm{KO,gauge}}$ to drop to zero at punctures by multiplying by the conformal factor $W$ raised to some fractional power.

While this refinement was quite effective in the context of the BBH evolution presented in that paper, $\epsilon_{\mathrm{KO,gauge}}(t,r,W)$ would need to be modified and potentially retuned for BBH scenarios at different initial separations and potentially different AMR grid structures.

Here we introduce a much simpler adjustment to $\epsilon_{\mathrm{KO}}$, which we call ``\cako'': Curvature-Adjusted KO dissipation. It multiplies the KO dissipation strength by the conformal factor $W=e^{-2 \phi}$, which ensures the KO strength is nearly zero near punctures, tending to $\epsilon_{\rm KO,CA}$ in the weak field:
\beq
\epsilon_{\rm KO}(W) = W \epsilon_{\rm KO,CA}.
\eeq
Similar to our past work, we set $\epsilon_{\rm KO,CA}$ to $0.99$ (near the CFL-limited maximum) for gauge quantities and to $0.3$ for all other BSSN evolved variables. But unlike our past work, this prescription has been validated using both \bhah grid structures across more than 100 unique scenarios across BBH parameter space, and (as shown in this work) AMR grid structures within the \et in the context of a GW150914-like BBH system at multiple numerical resolutions.

\subsection{\texorpdfstring{Coarse-grid-Adjusted $\mathcal{H}$ Damping (\cahd)}{Coarse-grid-Adjusted Hamiltonian-Constraint Damping}}
\label{subsec:Hconstdamping}

In 2002,~\cite{Yoneda:2002kg} found that analytically, the following modification to the right-hand side (RHS) of the conformal factor evolution equation causes one constraint-violating mode to exponentially damp:
\beq
\partial_t \phi = [\text{\eqref{eq:phievol} RHS}] + \kappa_{\phi\mathcal{H}}\alpha\mathcal{H}_{-} ,
\eeq
where $\kappa_{\phi\mathcal{H}} \le 0$ for stability.

As $\mathcal{H}=0$ in the continuum limit, this modification drops to zero with increasing resolution.  Further, for unit consistency $\kappa_{\phi\mathcal{H}}$ must have units of mass $M$ (i.e., the same as units of time and space in $G = c = 1$ units).

When analyzing the principal part of the adjusted $\partial_t \phi$ equation in the weak-field limit, we find
\beq
\partial_t \phi = \mathcal{H}_{-, \rm pp} = -  8 \kappa_{\phi{\cal H}} \partial_i \partial^i \phi,
\label{eq:ppart}
\eeq
which is a {\it parabolic} PDE. If this parabolic equation is solved explicitly with a 3D Laplacian~\cite{Raithel:2022san,Etienne:2014tia} in flat space, then at a numerical grid point with $\Delta s=\min (\Delta x^i)$ across all three spatial dimensions $i$, the CFL-constrained timestep $\Delta t$ is given by
\beq
-8 \kappa_{\phi{\cal H}} \frac{\Delta t}{(\Delta s)^2} \le \frac{1}{6} \implies
\Delta t \le \frac{(\Delta s)^2}{-48 \kappa_{\phi{\cal H}}}.
\eeq
The $(\Delta s)^2$ on the right-hand side is undesirable. For instance, suppose a simulation is performed at the maximum stable $\kappa_{\phi{\cal H}}$ at a given resolution $\Delta s$ and corresponding CFL-limited $\Delta t \propto \Delta s$. If a higher resolution (smaller $\Delta s$) is found necessary for a convergence study, $\kappa_{\phi{\cal H}}$ would need to be adjusted downward in order for the CFL condition to be satisfied. This would result in an inconsistency in the strength parameter across different resolutions, potentially invalidating the convergence study.

To address this, \cite{Duez:2002bn} included the simulation timestep in the prefactor:
\beq
\partial_t \phi = [\text{\eqref{eq:phievol} RHS}] + K \Delta t \mathcal{H}_{+},
\label{eq:Duezetal}
\eeq
where $K$ is dimensionless and $K \ge 0$ for stability. Numerically, the term $\Delta t \propto \Delta s$ decreases the strength of the parabolic operator in direct proportion to the CFL limit, making this modification stable at all choices of grid resolution provided it is stable at a given choice of $K$. From an analytical perspective, both the $\Delta t$ prefactor and $\mathcal{H}$ drop to zero in the continuum limit. Choosing $K = 0.02$--$0.06$, they showed that this damping term reduced constraint violations by almost an order of magnitude in the context of BNS simulations.

More recently, \cite{Raithel:2022san} explored the effect of three different coefficient prescriptions in the context of BNS evolutions. These include: \eqref{eq:Duezetal} with $K$ set to a constant across all AMR grids; \eqref{eq:Duezetal} with $K$ \textit{decreased} on coarser grids according to $\Delta s_{\mathrm{min}}/\Delta s_n$ (where $\Delta s_n$ is the $\Delta s$ on the $n$th refinement level), and the third method,
\beq
\label{eq:prescriptionthree}
\partial_t \phi = [{\rm \eqref{eq:phievol}\ RHS}] + c_{\cal H} {\cal H}_+, \\
\eeq
where $c_{\mathcal{H}}$ is set to a fixed dimensionful constant (independent of $\Delta t$). They found the third approach preferable because it improved the convergence properties of the post-merger phase. Additionally, they attributed the decreasing effectiveness of the \cite{Duez:2002bn} approach at higher resolutions to its relatively poor convergence behavior. As discussed, setting the coefficient on the parabolic operator to a fixed value results in instability if the evolution is repeated with sufficiently high resolution, so this approach is best applied when the highest resolution can be known \textit{a priori}.

Our improvement to the approach inverts the second prescription of~\cite{Raithel:2022san}. Instead of decreasing the strength $K$ on coarser grids, we increase the strength according to the maximum allowed timestep on the refinement level, similar to the approach of~\cite{Assumpcao:2021fhq} for accelerating hyperbolic relaxation waves across a curvilinear numerical grid. Specifically, we adjust \eqref{eq:Duezetal} as follows:

\beq
\partial_t \phi = [\text{\eqref{eq:phievol} RHS}] - C \Delta t_n \left[\text{CFL}_0 \frac{\Delta s_n}{\Delta t_n}\right] \mathcal{H}_{-},
\eeq
where the $n$ subscript refers to the local value of a given quantity on grid $n$, and $\text{CFL}_0$ is the CFL factor on numerical grid $n=0$, which contains the smallest timestep. We find $C=0.15$ results in stable evolutions across BBH parameter space explored by \bhah, including nonzero spins and mass ratios from $q=1$ to $q=6$, and we generally set $\text{CFL}_0$ to 90\% of its highest possible value as a safety margin.

On the finest AMR level ($n=0$), the bracketed term reduces to 1. This approach maximally benefits standard (2:1 refined) AMR grids with a global timestep $\Delta t_n=\Delta t_0=\Delta t$, as it \textit{exponentially} increases the damping strength with $n$; in this limit the bracketed term reduces to $2^n$. For the $\Delta t_n$ used in this work (Sec.~\ref{sec:alg_approach}), the bracketed term increases the damping strength by a factor of $2^5=32$ on the coarsest grid.

\subsection{Slow-Start Lapse}
\label{subsec:SSL}

Our third and final improvement focuses on rapid changes in the lapse near puncture BHs at the start of a spacetime evolution. As illustrated in Fig.~\ref{fig:sharplapse}, these changes send a sharp pulse outward at $\sqrt{2}c$. To minimize the early dynamics, the initial lapse $\alpha_0$ is typically set to a ``pre-collapsed'' value of $\alpha_0=W_0$, where $W_0=e^{-2\phi_0}$ is the conformal factor $W$ at $t=0$.

Our slow-start lapse (\ssl) technique modifies the 1+log lapse evolution condition \eqref{eq:onepluslog} as follows:

\beq
\partial_t \alpha = [\text{\eqref{eq:onepluslog} RHS}] - W \left[h e^{-t^2 / (2 \sigma^2)}\right] (\alpha - W).
\eeq

\ssl consists of three terms multiplied together. First, $-(\alpha-W)$ exponentially damps $\alpha$ to the (evolved) conformal factor $W$. By convention, at $t=0$, $\alpha$ is set to $W$, rendering this term exactly zero at $t=0$. Second, the Gaussian prefactor in square brackets sets the inverse damping timescale, such that the damping timescale is short and effective early in the evolution but grows to a very large and inert value later on. Third, to ensure that the wormhole-to-trumpet~\cite{Brown:2009ki} transition inside the horizon is not affected, the damping strength is multiplied by an overall prefactor of $W$, which drops to zero at the puncture and smoothly increases away from the puncture.

In short, the \ssl modification acts to quench rapid deviations of $\alpha$ from the quantity to which it is initially set, $W$, just outside the horizon, arresting the development of the sharp lapse feature in this region.

After many experiments across $\sim$100 unique scenarios in BBH parameter space with \bhah, we find that a Gaussian height of $h = \frac{3}{5}M$ and a standard deviation of $\sigma = 20.0M$ are particularly effective in mitigating the initial lapse pulse. With these choices, \ssl damps on a timescale of $e^{t^2 / (2\sigma^2)} / h = \frac{5}{3}M$ at $t = 0$, ${\sim}10^2M$ at $t = 57M$, ${\sim}10^{5}M$ at $t = 94M$, and ${\sim}10^{16}M$ at $t = 170M$. In this way, \ssl impacts only the formation of the sharp lapse feature and is rendered completely ineffective shortly thereafter.

\section{Algorithmic Approach}
\label{sec:alg_approach}

We present reproductions of the GW150914 gallery example~\cite{wardell_barry_2016_155394} from the \et~\cite{EinsteinToolkit,EinsteinToolkit:web}, incorporating the aforementioned improvements in a systematic manner. This gallery example is notable, as it is perhaps the most comprehensively documented BBH NR simulation available; all NR codes, results, and plotting scripts required to reproduce the original simulation are openly accessible on Zenodo~\cite{wardell_barry_2016_155394}.

The physical parameters of the simulation are inspired by the first-ever directly detected GW event, GW150914~\cite{Abbott:2016blz}. In particular, it uses the best-fit parameters for GW150914~\cite{TheLIGOScientific:2016wfe}: a mass ratio of $q=\frac{36}{29}$, with dimensionless spin parameters of $\vec{\chi}_M=\{\chi_M^x,\chi_M^y,\chi_M^z\} = \{0,0,0.31\}$ for the more-massive BH and $\vec{\chi}_m=\{\chi_m^x,\chi_m^y,\chi_m^z\} = \{0,0,-0.46\}$ for the less-massive one. The initial separation is set to $10M$, which places the BHs approximately 5 orbits before common-horizon formation.

As in the gallery example, we use the \et~\cite{EinsteinToolkit,EinsteinToolkit:web} and its \Carpet~\cite{Carpet,CarpetCode:web} AMR infrastructure to perform these simulations. The numerical parameters for the evolution are largely the same: 8th-order finite differencing, 5th-order prolongation in space, 2nd-order prolongation in time, RK4 timestepping. The \TwoPunctures thorn is used to construct initial data \cite{AnsorgTwoPunctures}, \ahfd~\cite{AHFinderDirect} for apparent-horizon diagnostics, \qlm~\cite{Dreyer:2002mx} for isolated horizon diagnostics, and the \kranc~\cite{Husa:2004ip}-generated \weylscal for $\psi_4$ extraction. However, there are some notable differences.

First, instead of the \texttt{ML\_BSSN}~\cite{Brown:2008sb,Husa:2004ip,McLachlan:web} thorn, the \nrpy-generated~\cite{Ruchlin:2017com} \BaikalVacuum thorn, modified as needed with the improved techniques, is used for the spacetime evolution. \BaikalVacuum and its sister code \Baikal are BSSN evolution codes in the \et: \BaikalVacuum solves the BSSN equations without stress-energy source terms, and \Baikal includes these source terms. The latter was used in~\cite{Armengol:2021mbt} and~\cite{Werneck:2022exo} for the evolving the spacetime fields in BNS simulations.

Second, the \TwoPunctures grid resolution was increased from its default values of $N_A \!\times\! N_B \!\times\! N_\phi = 30 \!\times\! 30 \!\times\! 16$  to $48 \!\times\! 48 \!\times\!20$, as the default values result in initial data constraint violations dominating the early evolution. Also, the initial lapse was chosen to be $W=e^{-2\phi}$, instead of the ``\verb|TwoPunctures-averaged|'' value, which sets the lapse to $\alpha(r) = \frac{1}{2} \sum_{i=1}^2 (1-m_i/(2 r_i))$ for each puncture $i$ of mass $m_i$ at distance $r$ from $r_i$. The former approach has the advantage of exhibiting the same asymptotic behavior at large radius and late times: $\alpha_{\infty} = 1 - M/r$ for total mass $M$. Meanwhile, the latter approach tends to $1 - M/(2r)$ in the weak field. While \TwoPunctures has an option to set the initial lapse to any power of $\psi=e^{\phi}$, it chooses a value of $\psi$ that does not include the correction term $u$ (Eq.~30 in \cite{AnsorgTwoPunctures}). To set the initial value of the lapse to the evolved quantity $W$, as needed for \ssl, we modified \TwoPunctures to include this option. Our modified version may be found in this GitHub repository~\cite{BaikalVacuum2024}.

Third, the \Llama~\cite{LlamaCode} cubed-spheres grid in the gravitational wave zone was replaced with box-in-box Cartesian adaptive-mesh refinement (AMR) grids of comparable or higher resolution. In addition, we choose a default resolution of all AMR grids 50\% higher, so that, for example, the resolution at the punctures is $M/42$ instead of the gallery example value of $M/28$. As we will find in Sec.~\ref{sec:results}, even $M/42$ lies slightly outside the convergent regime for modeling this physical scenario. For completeness, our AMR grid hierarchy comprises 11 levels of overlapping grid cubes, centered on and comoving with each puncture, with a standard factor-of-2 refinement across adjacent AMR levels (i.e., there are 10 levels of refinement built upon the coarsest level). The AMR grid cubes have half-sidelengths of $\left\{\frac{2}{3}M, \frac{4}{3}M, \frac{8}{3}M, \frac{16}{3}M, \frac{64}{3}M, \frac{128}{3}M, ..., \frac{4096}{3}M\right\}$. As our default choice we choose grid resolutions of $\left\{\frac{M}{42}, \frac{M}{21}, \frac{M}{10.5}, ..., 24.38M\right\}$ with CFL factors of $\frac{1}{640}\{288, 288, 144, 72, 36, 18, 9, 9, 9, 9, 9\}$, respectively.

Finally, the imposed symmetry across the $z$-axis was removed as \BaikalVacuum does not currently support symmetries.

\section{Results}
\label{sec:results}

To investigate the individual and combined effectiveness of the improvements introduced in this work, we conducted five simulations based on the GW150914-inspired scenario detailed in Sec.~\ref{sec:alg_approach} at the default resolution. These simulations are denoted as \control, \cako, \cahd, \ssl, and \allthree. \control serves as the no-improvement baseline, while the \cako, \cahd, and \ssl each integrate a single enhancement: curvature-adjusted Kreiss-Oliger, $\mathcal{H}$-damping, and slow-start lapse, respectively. \allthree combines all three enhancements. For convergence analyses, we executed the \control and \allthree simulations at resolutions 1.25x, 1.5x, and 1.71x higher than the default resolution (1.0x)~\footnote{AMR grid structures and resolutions were chosen such that each resolution maintains consistent physical boundaries for AMR grids in our moving-box AMR algorithm.}.

\subsection{Effect of Improvements on Trajectories and Lapse Evolution}

\begin{figure}[ht!]
  \begin{center}
    \vspace{-0cm}
    \includegraphics*[angle=0,width=0.48\textwidth]{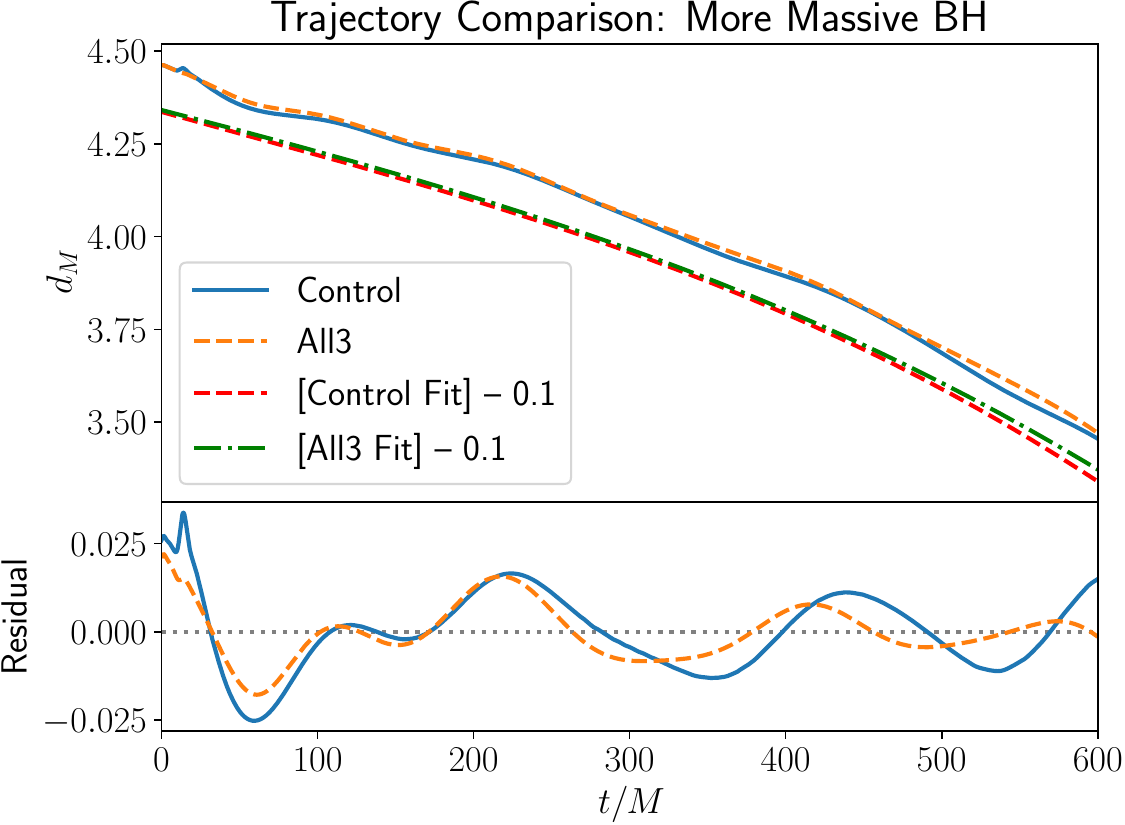}
    \includegraphics*[angle=0,width=0.48\textwidth]{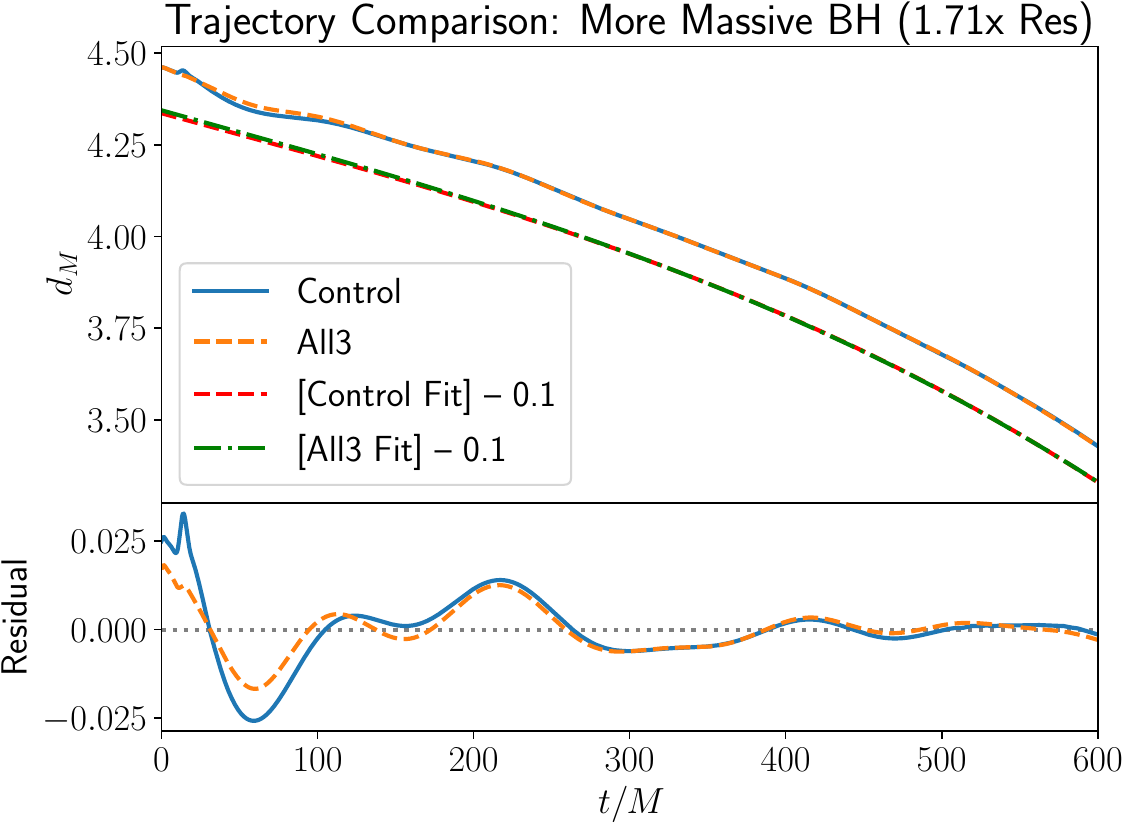}
  \end{center}
  \vspace{-0.4cm}
  \captionsetup{width=0.5\textwidth}
  \caption{\textbf{Trajectory comparison: \control vs. \allthree}. \textbf{Top panels, upper curves}: distance from the origin of the more massive BH $d_M$, comparing \control (blue solid) and \allthree (orange dashed) over the first three orbits. \textbf{Top panels, lower curves}: least-squares cubic polynomial fit $P_3(t)$ to \allthree (red dotted) and \control (green dotted), minus 0.1 to improve visibility. \textbf{Bottom panels}: difference between data and cubic polynomial fit.}
  \label{fig:trajectories}
\end{figure}

As the three improvements modify the evolution equations in the finite-resolution limit, as well as the 1+log lapse condition, it is reasonable to be concerned that \allthree may influence the trajectories of the BHs, especially at lower resolutions, possibly inducing unwanted eccentricity. To investigate this, in Fig.~\ref{fig:trajectories} a cubic polynomial $P_3(t)$ is fit to $d_M(t)$---the distance of the more massive BH from the center of mass (the origin)---over the first three orbits ($0\le t \le 600M$).

$d_M(t)$ and $P_3(t)-0.1$ are plotted in the top panels of each plot. Comparing the top and bottom plots, the $P_3(t)$ fits---and thus the trajectories---appear to converge with increased resolution. In addition, the sharp non-convergent oscillation in \control at $t \approx 20M$ is eliminated in \allthree. Comparing the effects of the individual improvements, we find that the \ssl enhancement is responsible for the removal of this sharp feature. As \ssl most significantly impacts the early gauge evolution, this is not surprising.

Regarding possible eccentricity enhancement in \allthree, the residual $d_M(t)-P_3(t)$ (bottom panels of Fig.~\ref{fig:trajectories}) shows the \textit{opposite} is the case: at both default and highest resolutions, the coordinate eccentricity over the first three orbits is slightly \textit{lowered} in \allthree as compared to \control. That is to say, the amplitude of the residual oscillation is reduced in \allthree as compared to \control. Further analysis reveals that \allthree reduces the average absolute value of the residual by 33\% at low resolution, from 0.0088 in \control to 0.0059 in \allthree; and by 25\% at the highest resolution, from 0.0059 to 0.0044. We stress that this is a measure of the smoothness of $d_M(t)$, not eccentricity.

We next turn our attention to further analyzing the effects of \ssl. As discussed in Sec.~\ref{subsec:SSL}, \ssl slows the deviation of the lapse away from $\alpha = W$ outside the horizon at early times. Given that we adopt the standard ``pre-collapsed'' lapse choice $\alpha = W$ at $t = 0$, \ssl has no impact on the evolution at $t = 0$.

\begin{figure}[ht!]
  \begin{center}
    \vspace{-0cm}
    \includegraphics*[angle=0,width=0.48\textwidth]{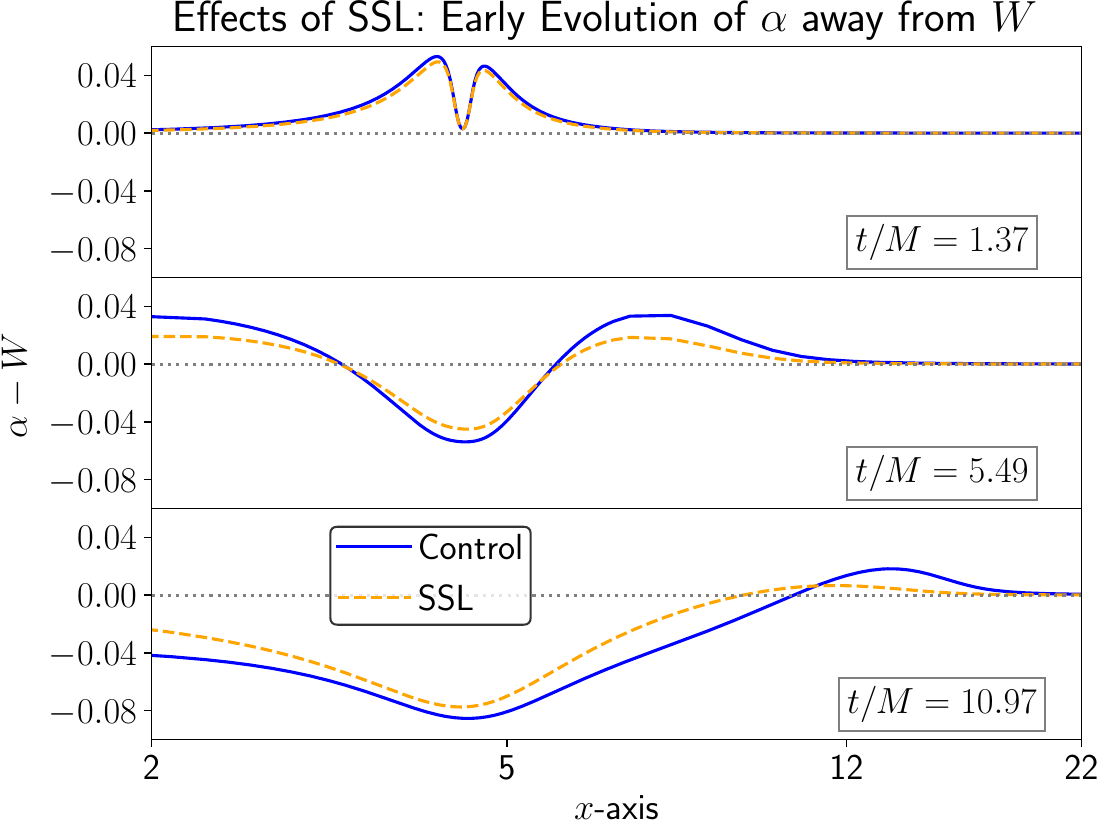}
  \end{center}
  \vspace{-0.4cm}
  \captionsetup{width=0.5\textwidth}
  \caption{\textbf{Effects of \ssl on Early Lapse Evolution}: $\alpha-W$ at various times during the early evolution along the $x$-axis, in the vicinity of the more massive BH, comparing \control (blue solid) with \ssl (orange dashed).}
  \label{fig:early_lapse_evolution}
\end{figure}

Figure~\ref{fig:early_lapse_evolution} illustrates how the evolution of the lapse deviates from the conformal factor shortly after the simulation begins, comparing $\alpha - W$ in \control to \ssl at times $t/M=1.37$, 5.49, and 10.97. For comparison, the first orbital period completes at $t\approx 222M$.

The more massive puncture is initially situated at $x = 4.46$, moving off the $x$-axis shortly after the evolution begins. Thus the top panel shows the evolution very close to the puncture, and lower panels depict slices at larger distances from the puncture and at later times. The conformal factor along the axis smooths significantly away from the puncture, so the oscillations in $\alpha - W$ at $t \ge 5.49M$ are primarily due to the lapse. While the effects of \ssl are barely noticeable at $t = 1.37M$, they become quite apparent starting at $t = 5.49M$; \ssl causes the oscillations in the lapse to be greatly reduced in amplitude, both at $t \approx 5.49M$ and $t \approx 10.97$.

\begin{figure}[ht!]
  \begin{center}
    \vspace{-0cm}
    \includegraphics*[angle=0,width=0.48\textwidth]{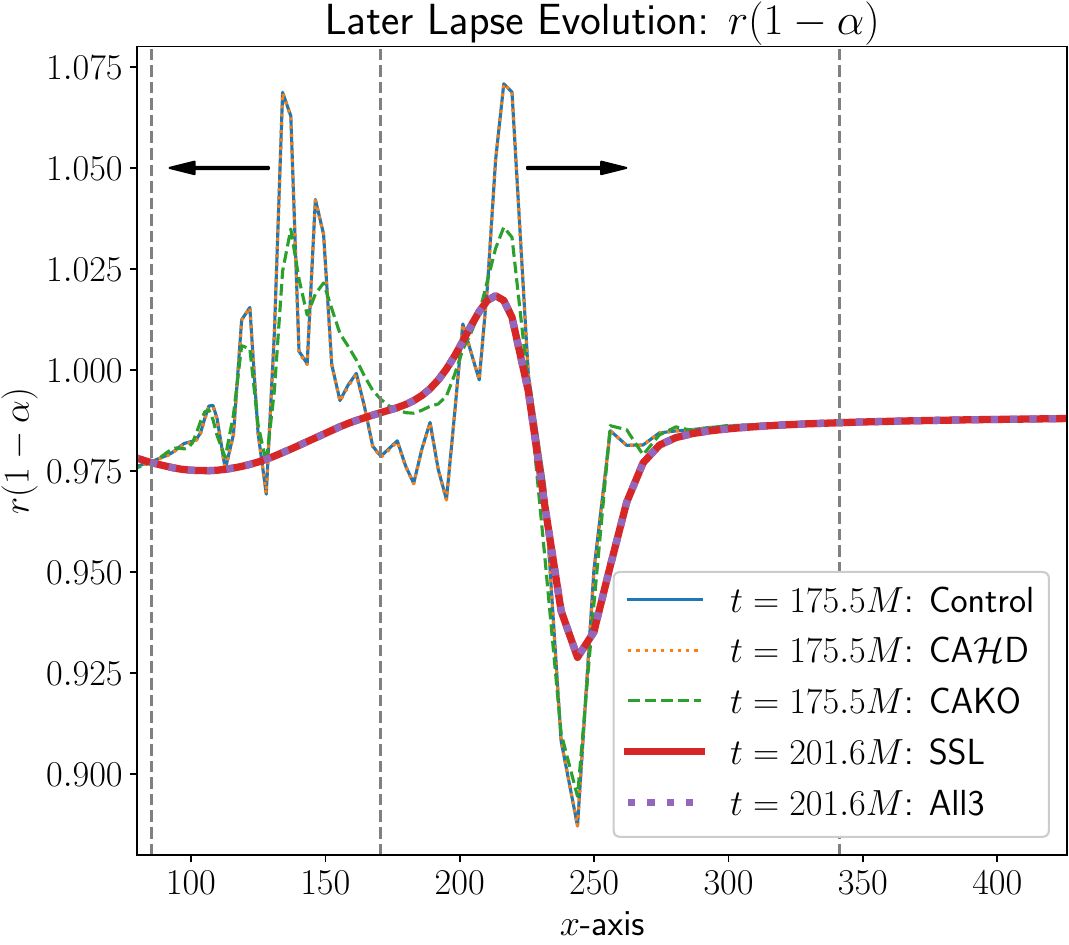}
  \end{center}
  \vspace{-0.4cm}
  \captionsetup{width=0.5\textwidth}
  \caption{\textbf{Impact of improvements on the sharp lapse feature}: $r(1-\alpha)$ is plotted at $t/M=175.5$ along $x$-axis for \control (thin solid), \cahd (thin dotted; overlaps \control), and \cako (thin dashed); and at $t/M=201.6$ for \ssl (thick solid) and \allthree (thick dotted; overlaps \ssl). Dashed vertical lines demarcate AMR boundaries along the $x$-axis at $x=85.\bar{3}$, $170.\bar{6}$, and $341.\bar{3}$. The transmitted and reflected wave pulse propagation directions in \control, \cahd, and \cako across the $170.\bar{6}$ AMR grid boundary are denoted by the horizontal arrows.}
  \label{fig:r_1malp_t175}
\end{figure}

The sharp lapse feature propagates outward and induces a great deal of constraint violation in its wake, as discussed in Sec.~\ref{sec:intro}. Figure~\ref{fig:r_1malp_t175} shows how each improvement impacts the sharp lapse feature evolution in the wavezone. As in Fig.~\ref{fig:sharplapse}, $r(1-\alpha)$ is plotted, which should tend to a constant value in the limit $r\to\infty$.

At $t=175.5M$, the wavefront in \control, \cahd, and \cako has propagated to roughly $x=250M$, consistent with a wave propagating from near the origin at $t=0$ at a speed of $\sqrt{2}c$. Notice this feature partially reflects off of the $x=170.\bar{6}M$ AMR grid boundary, due to undersampling across this factor-of-two drop in resolution. Meanwhile, the delayed evolution of the lapse away from $W$ early in the \ssl and \allthree evolution acts to delay the wavefront by approximately $t=26M$; these wavefronts at $t=201.6M$ overlap those of runs without \ssl at $t=175.5M$.

As the \cahd and \control results are indistinguishable, we conclude that \cahd has no significant impact on the evolution of the lapse itself. On the other hand, \cako acts as a strong low-pass filter in this weak-field region, significantly damping oscillations in the sharp lapse feature and its reflections over the standard KO prescription adopted in \control and \cahd.

Most notably, the \ssl modification to the 1+log lapse condition significantly smooths the outgoing lapse pulse, resulting in virtually no reflection across the $x=170.\bar{6}M$ grid boundary. While this grid boundary generates minimal noise in \ssl and \allthree (which perfectly overlap in this plot), partial reflection is observed in \ssl and \allthree when the wavefront crosses the $x=341.\bar{3}M$ grid boundary. After partially reflecting, \allthree exhibits less noise than \ssl due to the enhanced KO dissipation implemented in \cako.

\subsection{Strong-Field Region: Effect of Improvements on Constraint Violations}

\begin{figure}[ht!]
  \begin{center}
    \vspace{-0.cm}
    \includegraphics*[angle=0,width=0.48\textwidth]{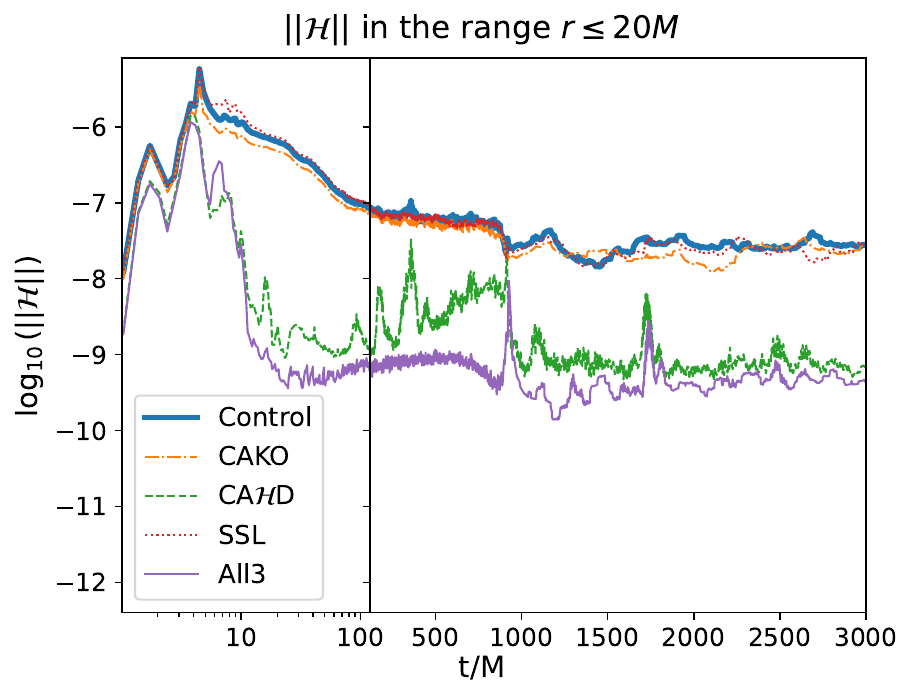} \\
    \includegraphics*[angle=0,width=0.48\textwidth]{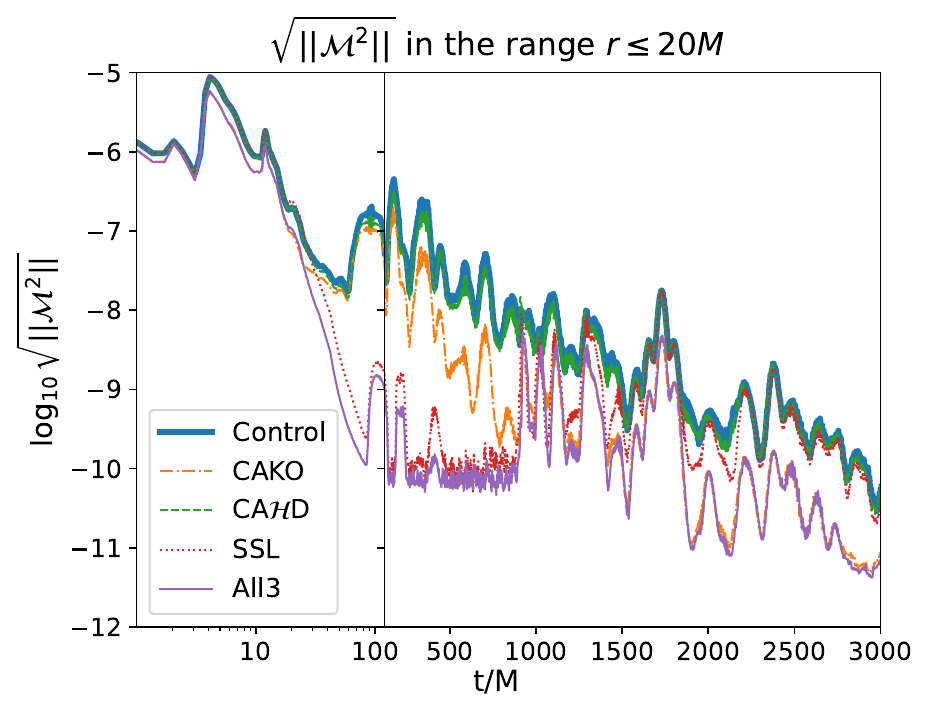}
  \end{center}
  \vspace{-0.4cm}
  \captionsetup{width=0.5\textwidth}
  \caption{\textbf{Effect of improvements: Constraint violation in strong-field region}: Volume-averaged $L_2$ norm of $||\mathcal{H}||$ (\textbf{top plot}) and $\sqrt{||\mathcal{M}^2||}$ (\textbf{bottom plot}) inside spherical region of radius $20M$, centered at the origin (center of mass), excluding two spheres of radius $1M$ centered at each BH. Results from five runs are compared: \control (thick solid blue), \cako (dash-dotted orange), \cahd (dashed green), \ssl (dotted red), and \allthree (thin solid purple). Volume excludes spheres of radius $r=1M$ centered at each BH.}
  \label{fig:strongfield_constraints}
\end{figure}

As discussed in~\cite{Etienne:2014tia} and illustrated in Fig.~\ref{fig:sharplapse}, the outward-propagating sharp lapse pulse fills the computational domain with significant constraint violations in its wake. The top plot in Fig.~\ref{fig:strongfield_constraints} compares how our improvements, both individually and collectively, reduce the volume-averaged $L_2$ norm of the Hamiltonian-constraint violation $||\mathcal{H}||$ in the strong-field region.

\cahd has the most significant influence on $||\mathcal{H}||$, reducing it by between 2--3 orders of magnitude after $t=10M$. This is consistent with its intended effect of causing one Hamiltonian-constraint-violating mode to exponentially damp. Further, \allthree reduces $||\mathcal{H}||$ in the strong field by an additional 0.2--1.2 orders of magnitude. The spike in constraint violation observed at $t\approx 930M$ is coincident with common horizon formation. While the same spike is only observed in \cahd and \allthree, the Hamiltonian-constraint violation in these cases never exceeds \control, \cako, or \ssl.

Meanwhile, as shown in the bottom plot of Fig.~\ref{fig:strongfield_constraints}, at most times \ssl greatly reduces momentum-constraint violation $\sqrt{||\mathcal{M}^2||}$ in the strong field at early times, decreasing it by between 0--3.5 orders of magnitude. After the merger, \cako proves to be most effective at reducing momentum-constraint violation. Once more, the cumulative impact of all improvements (\allthree) surpasses that of any individual improvement.

Next we analyze the convergence behavior of the constraint violations. Assuming we are in the convergent regime (i.e., the region where truncation, and not roundoff or undersampling error dominates), we would expect the constraints to converge to zero between second and eighth order in our mixed-order scheme; second order corresponds to the order at which data are interpolated (prolongated) in time at AMR grid boundaries, and eighth order corresponds to the finite-differencing order for spatial derivatives. Witnessing inconsistent convergence order in such a complex system typically implies that some physical lengthscale is not being properly resolved (undersampling error).

Given that constraint violations converge to zero in the continuum limit, in the truncation-error-dominated regime, we should find that the constraint violation $C_i$ at resolution $\Delta x_i$ and convergence order $n$ should approximately obey the equation:
\beq
C_i = A (\Delta x_i)^n,
\label{eq:truncerror}
\eeq
where $A$ is related to the $(n+1)$st derivative of the term in $C_i$ that dominates the truncation error. So long as $C_i$ is known at at least two resolutions, both $n$ and $A$ can be computed.

\begin{figure}[ht!]
  \begin{center}
    \vspace{-0.cm}
    \includegraphics*[angle=0,width=0.48\textwidth]{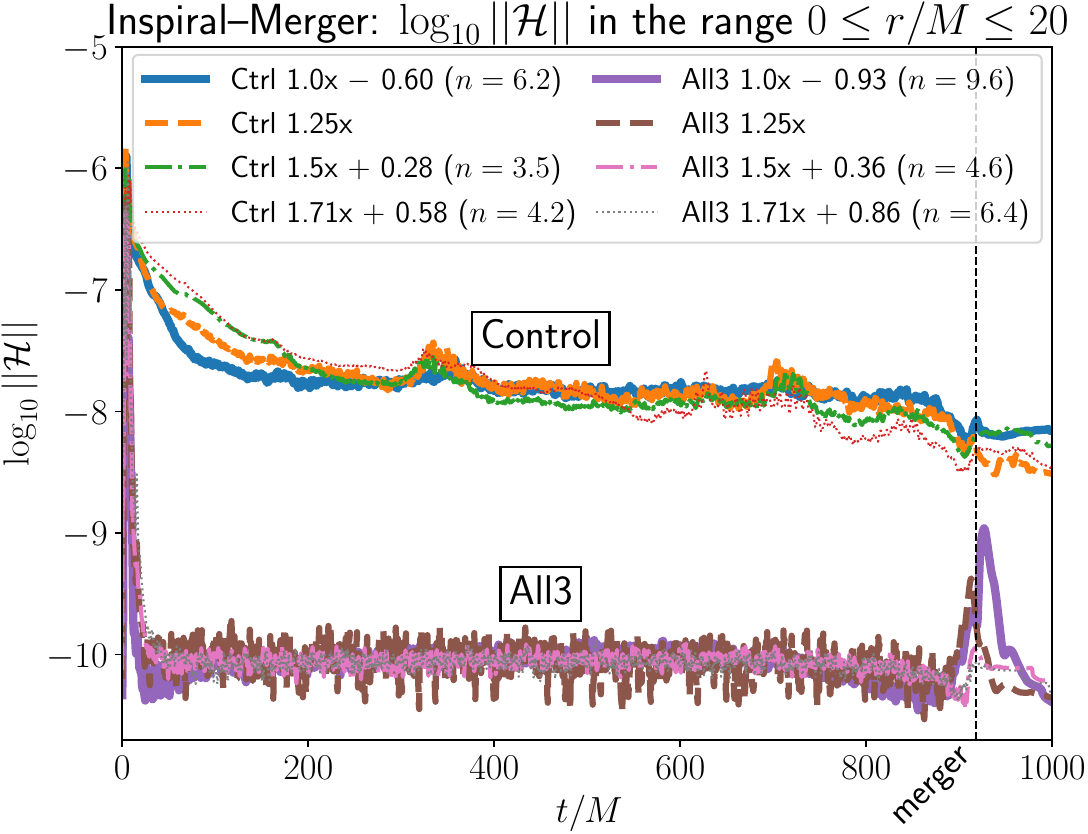} \\
    \includegraphics*[angle=0,width=0.48\textwidth]{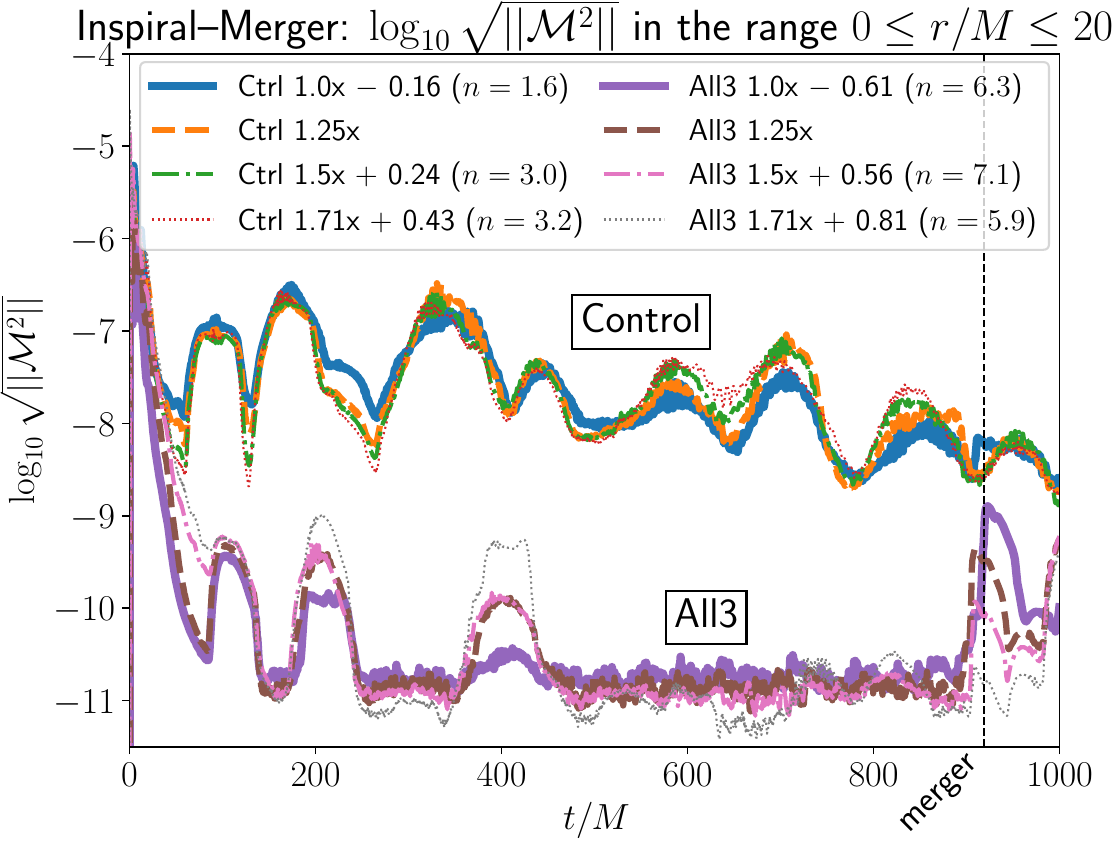}
  \end{center}
  \vspace{-0.4cm}
  \caption{\textbf{Convergence of constraint violation in strong-field region}: Convergence to zero of volume-averaged $L_2$ norms of Hamiltonian (\textbf{top plot}) and momentum (\textbf{bottom plot}) with increased resolution in a spherical region of radius $20M$, centered at the origin (center of mass), excluding two spheres of radius $1M$ centered at each BH. \control and \allthree results are compared. The lowest resolution, as well as 1.5x and 1.71x higher resolution results have been shifted up or down to maximally overlap the 1.25x higher resolution results. Shifts and implied convergence orders ($n$) are reported in the legend for each case.}
  \label{fig:strongfield_constraints_convergence}
\end{figure}

Figure~\ref{fig:strongfield_constraints_convergence} plots the base-10 logarithm $L_2$ norms of Hamiltonian ($||\mathcal{H}||$) and momentum ($\sqrt{||\mathcal{M}^2||}$) constraint violations in the strong-field region, through merger and ringdown, up to $t=1000M$. After this, the key payload from the simulation---the GW signal---has largely propagated away and no longer depends on data in this region. Data from \control and \allthree at the default (1.0x) resolution, as well as 1.5x and 1.71x higher resolutions are shifted on this logarithmic scale such that the difference with the corresponding 1.25x case is minimized; from this difference an implied convergence order $n$ can be immediately computed using Eq.~\ref{eq:truncerror}:
\beq
n = \frac{\log_{10} C_i - \log_{10} C_{1.25x}}{\log_{10} \Delta x_i - \log_{10} \Delta x_{1.25x}}.
\eeq

The top plot of Fig.~\ref{fig:strongfield_constraints_convergence} indicates that through most of the evolution, Hamiltonian-constraint violation in \allthree remains about \textit{100 times} lower than \control at 1.25x resolution. More strikingly, the implied convergence orders at higher resolutions are consistently higher in \allthree (4.6--6.4) compared to \control (3.5--4.2), indicating that this gap only widens at higher resolutions.

\allthree is even more effective at reducing momentum-constraint violations in the strong-field (bottom plot of Fig.~\ref{fig:strongfield_constraints_convergence}), dropping them by about 3--4 orders of magnitude as compared to \control at 1.25x resolution. Unlike the implied convergence order from Hamiltonian-constraint violations, the momentum constraint convergence order is nearly \textit{doubled} at higher resolutions in \allthree (5.9--7.1) compared to \control (3.0--3.2), reflecting that at higher resolutions, the 3--4 decade gap between \control and \allthree also widens. Finally, 1--2 order-of-magnitude bumps in $\sqrt{||\mathcal{M}^2||}$ appear periodically as reflections from distant AMR boundaries enter this region. These reflections strongly impact \allthree's convergence around $t/M=200$ and 400, but not at other times.

Analyzing implied convergence orders in Fig.~\ref{fig:strongfield_constraints_convergence}, two results stand out: \allthree's Hamiltonian-constraint violation exhibits $n=9.6$ (1.0x vs 1.25x) and \control's momentum-constraint violation exhibits $n=1.6$ (1.0x vs 1.25x). These implied convergence orders lie outside the expected range for our mixed-order scheme of $n=2$ to $n=8$, indicating that our chosen default resolution may be outside the convergent regime.

\begin{figure}[ht!]
  \begin{center}
    \vspace{-0.cm}
    \includegraphics*[angle=0,width=0.48\textwidth]{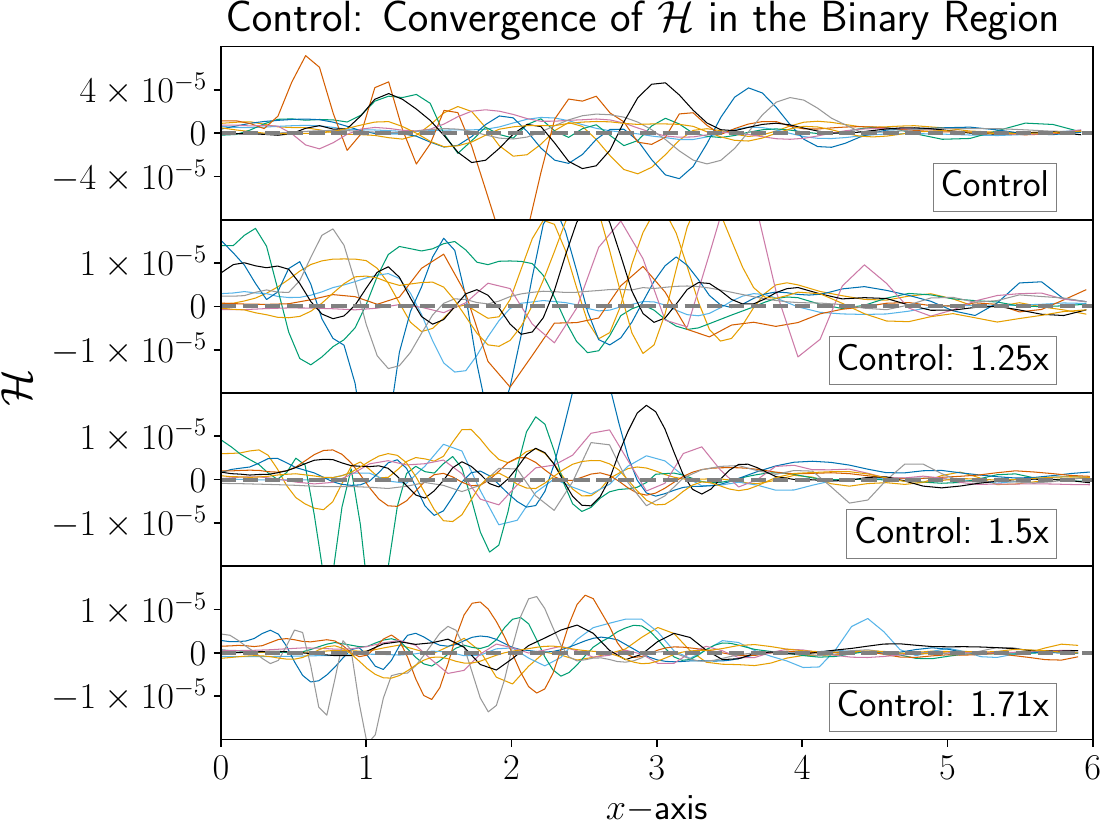} \\
    \includegraphics*[angle=0,width=0.48\textwidth]{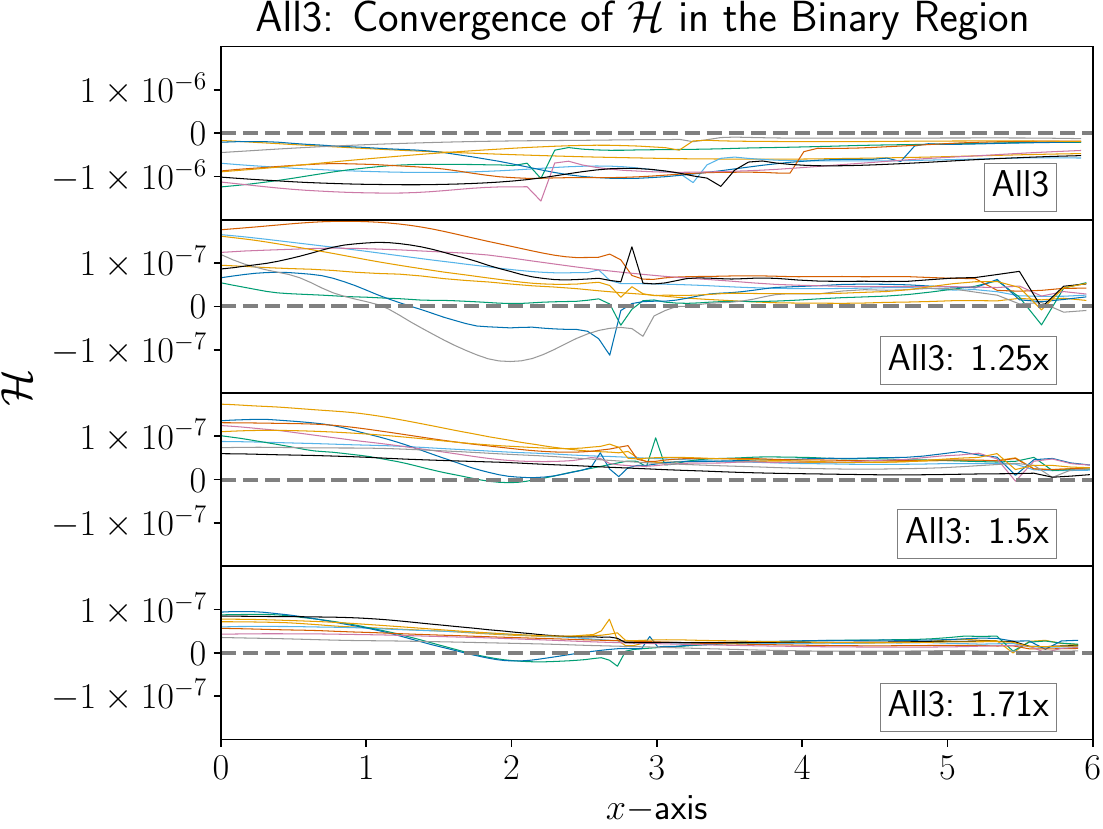}
  \end{center}
  \vspace{-0.4cm}
  \captionsetup{width=0.5\textwidth}
  \caption{\textbf{Hamiltonian-constraint violation in the binary region during inspiral}: Convergence of $\mathcal{H}$ (thin solid curves) in \control with increased resolution (\textbf{top plot}) and \allthree (\textbf{bottom plot}) along the positive $x$-axis at times when the BHs are very close to the $y$-axis, starting at approximately $0.75$ orbits: $t/M \approx 168$, $272$, $371$, $464$, $549$, $628$, $701$, $765$, and $828$. A thick, gray, dashed line denotes $y=0$. Note that both plots zoom out the top panel to better fit the data. The BHs are initially placed at $x = 4.46$ and $x=5.54$, and the origin is the center of mass. Merger occurs at roughly $t/M=930$.}
  \label{fig:binary_region_constraint_convergence}
\end{figure}

Unusual implied convergence orders warrant further investigation. In Fig.~\ref{fig:binary_region_constraint_convergence}, we examine Hamiltonian-constraint violation in the region where the binary orbits, along the $x$-axis at various times when the BHs are close to the $y$-axis. \control is characterized by sharp, wavelike features about $\mathcal{H}=0$, while \allthree exhibits a purely negative (at the lowest resolution) or mostly positive (at higher resolutions) systematic bias, albeit with constraint violation amplitudes roughly $1/100$th that of \control (consistent with Fig.~\ref{fig:strongfield_constraints_convergence}). For scale, \allthree data would appear nearly indistinguishable from the horizontal dashed lines marking $\mathcal{H}=0$ in the top plot.

While the amplitude of constraint violation indeed drops with increased resolution in both \control and \allthree, there is a clear zero crossing in \allthree when comparing lowest and 1.25x higher resolution cases, consistent with the lowest resolution not yet being in the convergent regime. Further, the 1.25x resolution does not exhibit a clear bias in \allthree, possibly implying that it lies very close to the boundary between the convergent and non-convergent regimes

As indicated by \eqref{eq:effectivedensity} and surrounding discussion, $\mathcal{H}$ violations act as an effective energy density $\rho_{\rm eff}$; thus, the consistent $\rho_{\rm eff} < 0$ in 1.0x \allthree implies that Hamiltonian-constraint violation acts to continually push the BHs away from one another. If this were the dominant error, we would expect a delayed merger. Following the same reasoning, the consistent $\rho_{\rm eff} > 0$ in 1.5x and 1.71x resolutions of \allthree imply an early merger, and the 1.25x resolution somewhere in-between. As will be discussed in our analysis of the extracted waveforms (Sec.~\ref{subsec:resultsweakfieldregion}), this is exactly what we observe. Such a clean analysis is impossible in \control due to the extremely high level of noise, and implies that clean waveform convergence in \control might depend on near-exact cancellation of $\mathcal{H}$ waves.

\begin{figure}[ht!]
  \begin{center}
    \vspace{-0.cm}
    \includegraphics*[angle=0,width=0.48\textwidth]{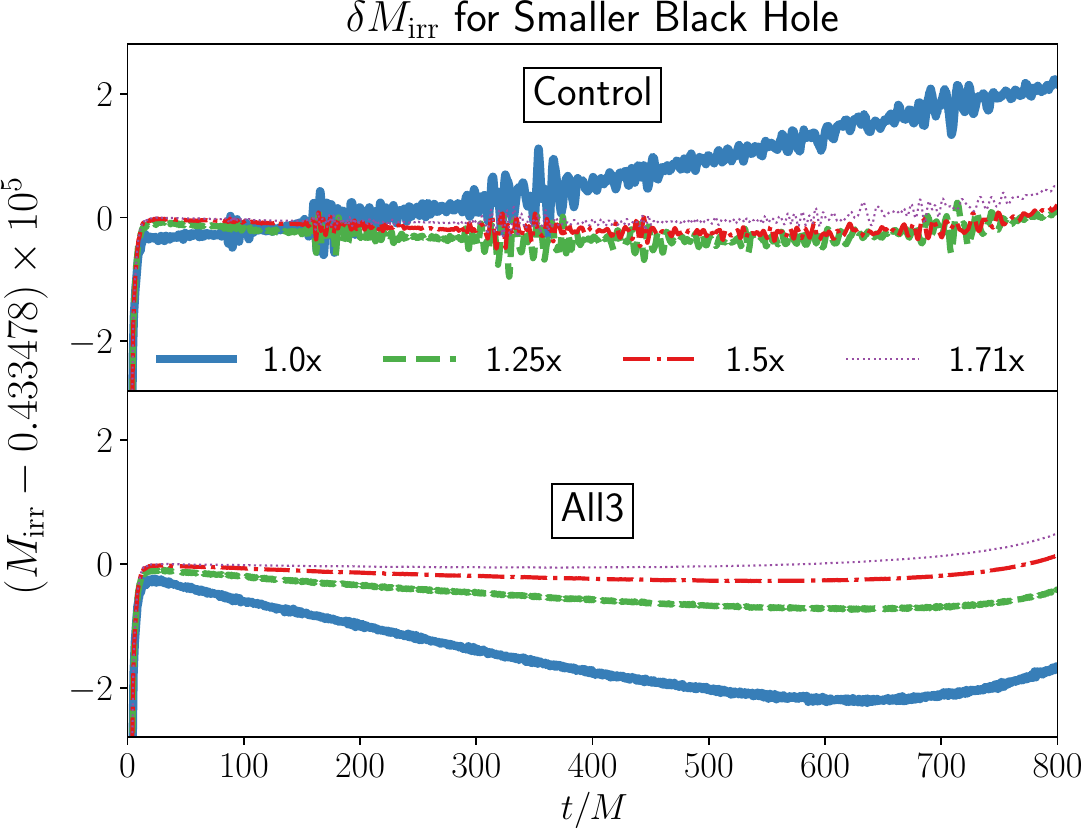}
  \end{center}
  \vspace{-0.4cm}
  \captionsetup{width=0.5\textwidth}
  \caption{\textbf{Convergence behavior of $M_{\rm irr}$, for smaller, faster-spinning BH}: Comparison of irreducible mass for \control (\textbf{top panel}) and \allthree (\textbf{bottom panel}) cases at all four resolutions. Thicker lines imply coarser resolution. Data beyond $t=800M$ are removed, as this is close to the time of common horizon formation; merger occurs at $t\approx 910M$.}
  \label{fig:irreducible_mass}
\end{figure}

We conclude that examining $\mathcal{H}$ in the binary region serves some use in identifying whether BBH evolutions are in the convergent regime, at least in the \allthree cases. Regarding \control we found a low implied convergence order of $||\mathcal{M}^2||$ when comparing 1.0x and 1.25x resolutions. Further analysis of the irreducible mass (Fig.~\ref{fig:irreducible_mass}) of the less massive, faster-spinning BH lends additional insight.

Figure~\ref{fig:irreducible_mass} shows that, after the early settling of $M_{\mathrm{irr}}$, the irreducible mass remains constant to within a couple of parts in $10^5$ across all cases. Comparing 1.0x with 1.25x and 1.5x resolutions, \control exhibits a \textit{negative} implied convergence order beyond $t \approx 150M$, providing further evidence that the 1.0x resolution data in the strong-field are not in the convergent regime. Measuring $n$ directly for \control is quite challenging due to numerical noise, though given the stochastic behavior observed in $\mathcal{H}$ near the punctures (Fig.~\ref{fig:binary_region_constraint_convergence}), this is unsurprising. That said, comparing 1.25x, 1.5x, and 1.71x resolution data yields an implied convergence order of roughly $n=1$ at $t=400M$, but the 1.25x and 1.5x data completely overlap at $t\approx 600M$ (implying $n=0$). Similar convergence behavior was observed in this quantity in~\cite{Etienne:2014tia}.

Meanwhile, \allthree exhibits far less noise and cleaner convergence in this diagnostic, with an implied convergence order of $n \approx 4.7$ in the range $200 < t/M < 600$ when analyzing the 1.0x, 1.25x, and 1.5x data, and $n \approx 2.0$ when analyzing the 1.25x, 1.5x, and 1.71x data. Although the \allthree data have implied convergence orders consistent with the range expected from our mixed-order scheme ($2 \le n \le 8$), Fig.~\ref{fig:binary_region_constraint_convergence} has established that constraint violations in this region in the 1.0x \allthree simulation are not in the convergent regime.

\subsection{Weak-Field Region}
\label{subsec:resultsweakfieldregion}

\begin{figure}[ht!]
  \begin{center}
    \vspace{-0.cm}
    \includegraphics*[angle=0,width=0.48\textwidth]{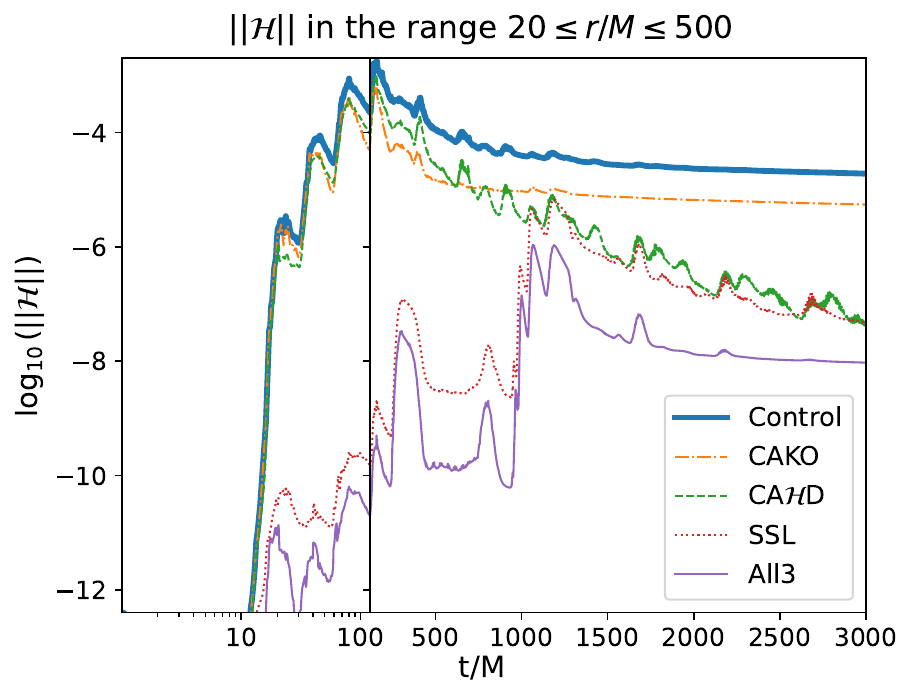}
    \includegraphics*[angle=0,width=0.48\textwidth]{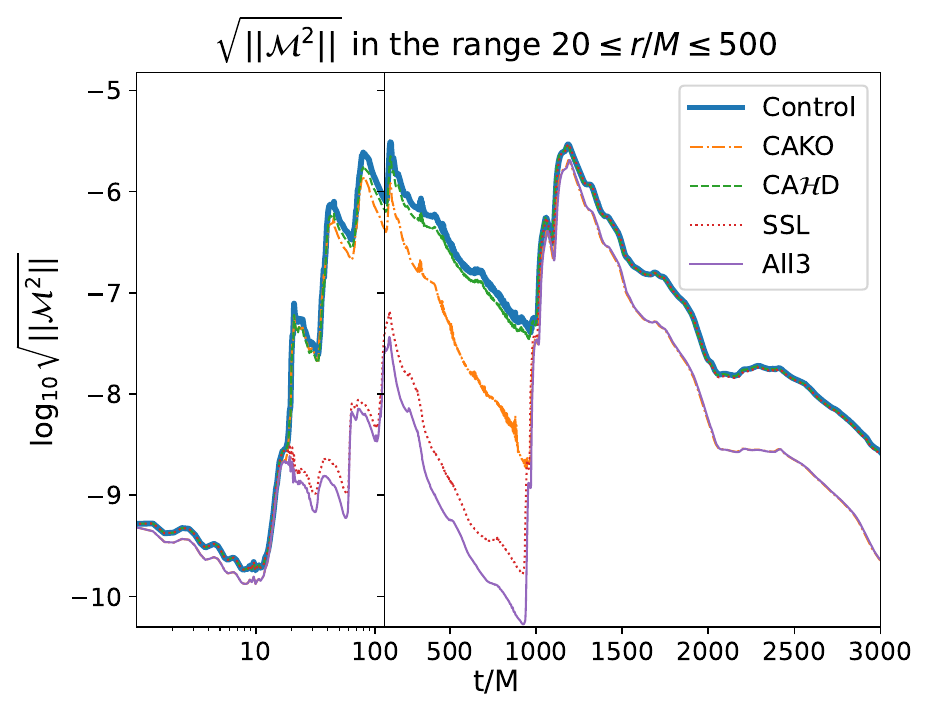}
  \end{center}
  \vspace{-0.4cm}
  \captionsetup{width=0.5\textwidth}
  \caption{\textbf{Effect of improvements: Constraint violation in gravitational wavezone}: Volume-averaged $L_2$ norm of $||\mathcal{H}||$ (\textbf{left}) and $\sqrt{||\mathcal{M}^2||}$ (\textbf{right}) within a thick spherical shell of $20\le r/M \le 500$, centered at the origin (center of mass). Results from five runs are compared: \control (thick solid blue), \cako (dash-dotted orange), \cahd (dashed green), \ssl (dotted red), and \allthree (thin solid purple).}
  \label{fig:weakfield_constraints}
\end{figure}

While numerical noise in \control significantly influences the constraints and $M_{\rm irr}$ in the strong field, it is most prominent in the weak field region where GW content---typically the core payload of BBH simulations---is extracted. As our improvements act to mitigate this noise, they are particularly effective in the weak-field region. Figure~\ref{fig:weakfield_constraints} demonstrates that throughout much of the inspiral ($t\lesssim 1000M$), Hamiltonian-constraint violations in the GW extraction region (the ``wavezone'') are reduced by roughly \textit{four to six} orders of magnitude in \allthree, largely a consequence of \ssl. This contrasts with the strong-field region (Fig.~\ref{fig:strongfield_constraints}), where 2--3 orders of magnitude reductions are observed. Meanwhile, during the inspiral, \allthree reduces momentum-constraint violations by between 2--3 orders of magnitude, similar to the strong-field region.

At the merger, a spike in Hamiltonian- and momentum-constraint violations propagates from the strong-field region into the wavezone. In \control, this spike is observed only in momentum-constraint violations, likely because noise levels in this region are far higher. Post-merger, Hamiltonian-constraint violation reductions in \allthree equilibrate to values roughly 3--4 orders of magnitude below \control, with near-equal contributions from \ssl and \cahd.

Careful analysis of momentum-constraint violations in this region after merger shows that \cahd and \ssl overlap \control, making \cako the dominant contributor to reducing momentum-constraint violations. While \allthree always falls below \control, they are quite comparable as constraint violations associated with common horizon formation propagate outward. Later, the gap between \control and \allthree slowly widens to about one order of magnitude by the end of the simulation.

\begin{figure}[ht!]
  \begin{center}
    \vspace{-0.cm}
    \includegraphics*[angle=0,width=0.48\textwidth]{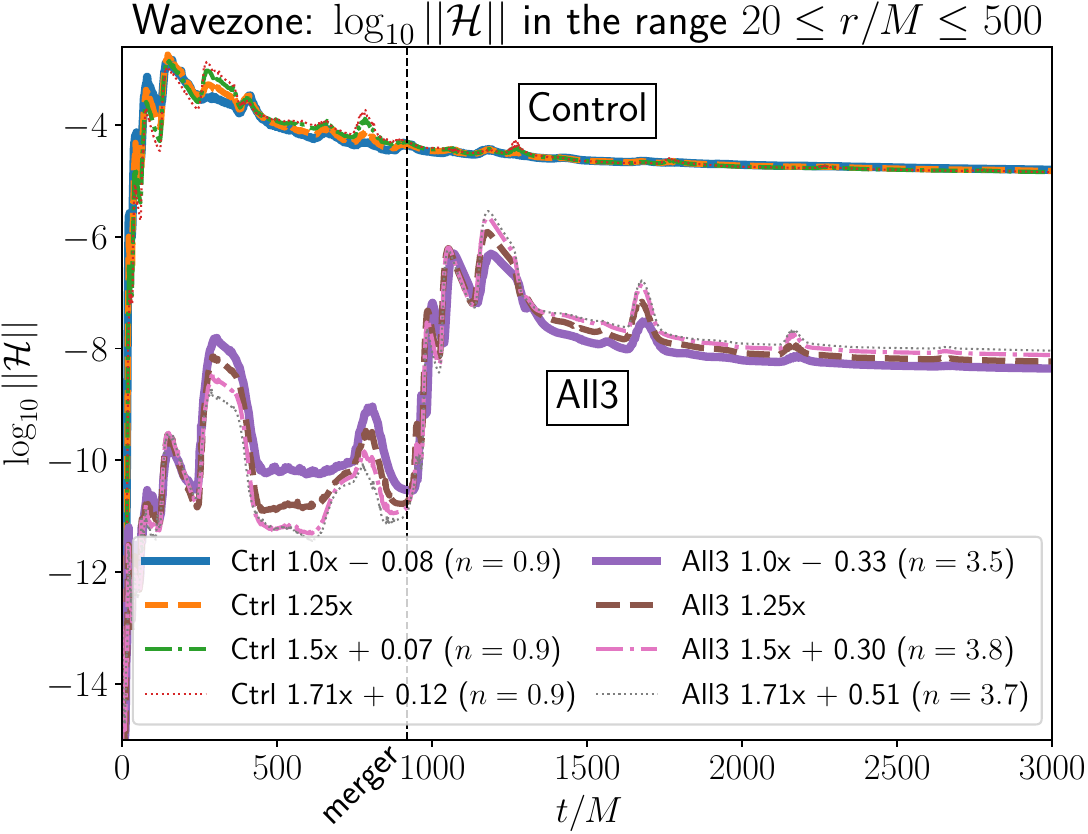} \\
    \includegraphics*[angle=0,width=0.48\textwidth]{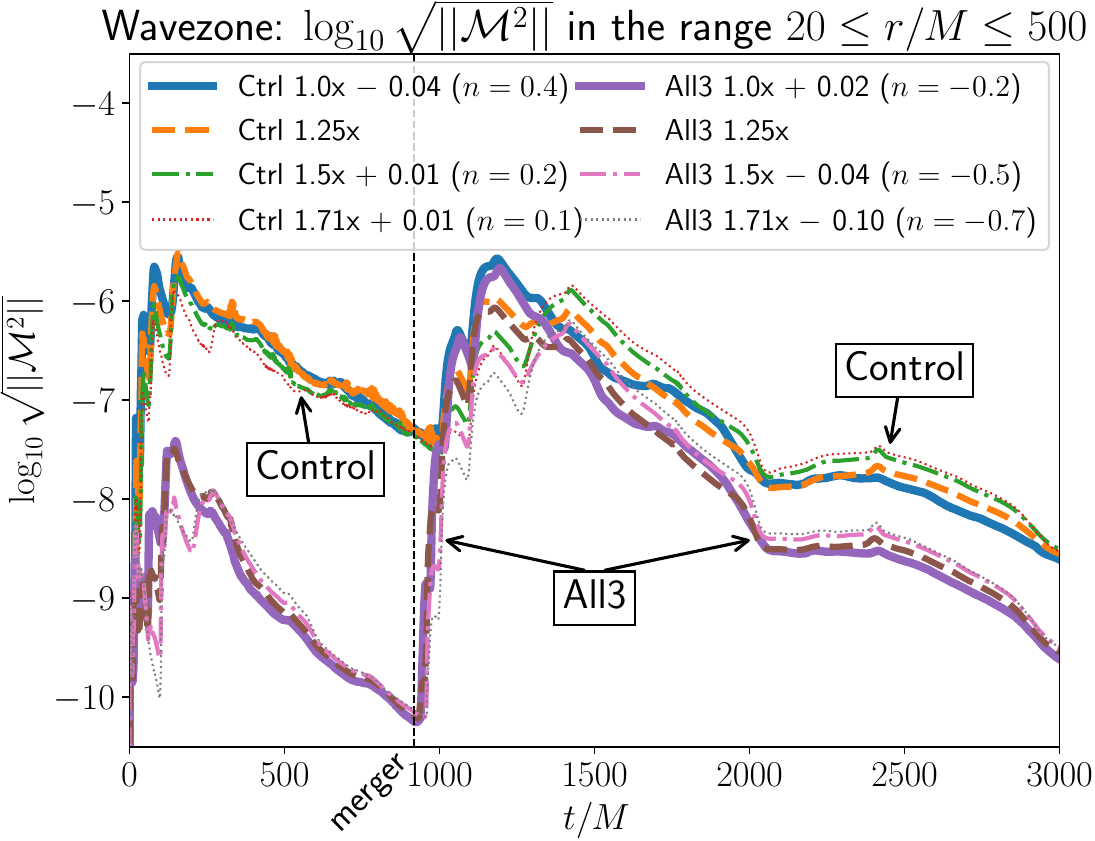}
  \end{center}
  \vspace{-0.4cm}
  \captionsetup{width=0.5\textwidth}
  \caption{\textbf{Convergence of constraint violation in gravitational wavezone}: Convergence to zero of volume-averaged $L_2$ norms of Hamiltonian (\textbf{top plot}) and momentum (\textbf{bottom plot}) with increased resolution in a thick spherical shell of inner radius $20M$ and outer radius $500M$, centered at the origin (center of mass). Lowest resolution as well as 1.5x and 1.71x higher resolution results have been shifted up or down to maximally overlap the 1.25x higher resolution results. Shifts and implied convergence orders ($n$) are reported for each case.}
  \label{fig:weakfield_constraint_convergence}
\end{figure}

Turning now to constraint convergence behavior in the wavezone at higher resolutions, Fig.~\ref{fig:weakfield_constraint_convergence} plots the constraint violation at all four resolutions, using the 1.25x resolution as the basis of comparison. In \control, neither Hamiltonian nor momentum-constraint violations converge to zero in the expected range of $2 \le n \le 8$, while, remarkably, \allthree exhibits implied convergence orders of 3.5--3.8 in the Hamiltonian-constraint violation. However, \allthree falls outside the expected convergence-order range in momentum-constraint violation, and in fact is slightly anti-convergent. While this is rather troubling for both \control and \allthree, upon closer analysis, we find that the level of violation in \allthree remains lower than \control at all resolutions, just as in Fig.~\ref{fig:strongfield_constraints_convergence}. We also find that if the integral is extended to the outer boundary as in~\cite{Etienne:2014tia}, we find a consistent result: both \control and \allthree exhibit convergence orders in the expected range $2 \le n \le 8$, implying convergent behavior outside of the wavezone.

Considering the GWs themselves, while GW detectors measure variation in GW strain, NR simulations typically compute the Weyl scalar $\psi_4$. As $\psi_4$ is the second time derivative of the strain in the weak-field limit---dependent on second spatial derivatives of the metric---it amplifies intrinsic numerical noise and thus provides an excellent noise diagnostic.

\begin{figure}[ht!]
  \begin{center}
    \vspace{-0.cm}
    \includegraphics*[angle=0,width=0.48\textwidth]{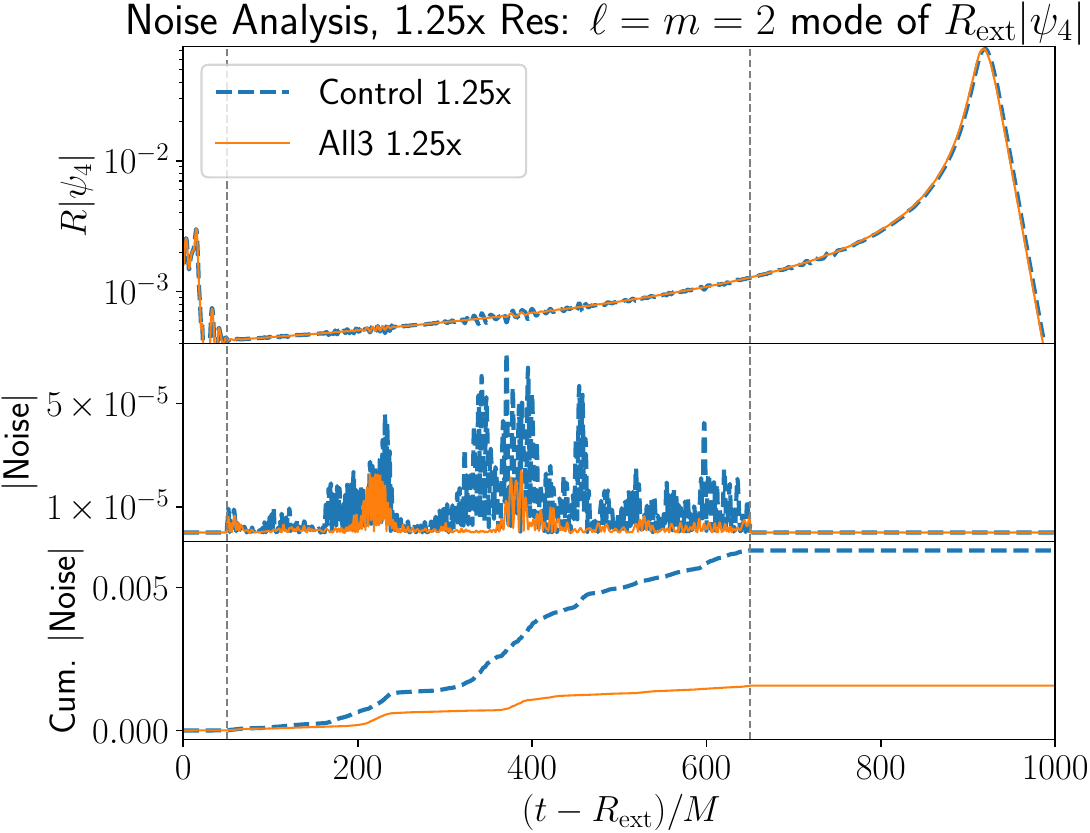}
  \end{center}
  \vspace{-0.4cm}
  \captionsetup{width=0.5\textwidth}
  \caption{\textbf{GW noise analysis: dominant mode}. \textbf{Top panel}: Amplitude of dominant ($\ell=m=2$) spin-weight $s=-2$ mode of $\psi_4$ versus retarded time. Vertical dashed lines: region where a quartic polynomial is fit to amplitude data. \textbf{Middle panel}: Absolute difference between data and fit. \textbf{Bottom panel}: Cumulative absolute difference between data and fit.}
  \label{fig:psi4noise_l2m2}
\end{figure}

Figure~\ref{fig:psi4noise_l2m2} illustrates our procedure for quantifying noise in the dominant, $\ell=m=2$, spin-weight $s=-2$ mode of $\psi_4$. In the top panel, we plot the amplitude of this mode extracted from the simulation at a radius $R_{\rm ext}=100M$, for both \control and \allthree at 1.25x higher resolution than the lowest resolution. The vertical dashed lines denote the region where a quartic polynomial was fit through the amplitude data. The middle panel plots the absolute difference between the data and the quartic polynomial fit (i.e., the ``noise''), and the bottom panel the noise integrated over time. Averaged over the time interval of the fit, our combined improvements reduce numerical noise in the $\ell=m=2$ mode of $\psi_4$ by a factor of 4.0 at 1.25x resolution. Applying the same technique to other resolutions, we find noise reduction factors of 6.0 (1.0x), 4.3 (1.25x), and 3.1 (1.71x). This underscores the fact that this noise converges away extremely slowly at higher resolutions.

\begin{figure}[ht!]
  \begin{center}
    \vspace{-0.cm}
    \includegraphics*[angle=0,width=0.48\textwidth]{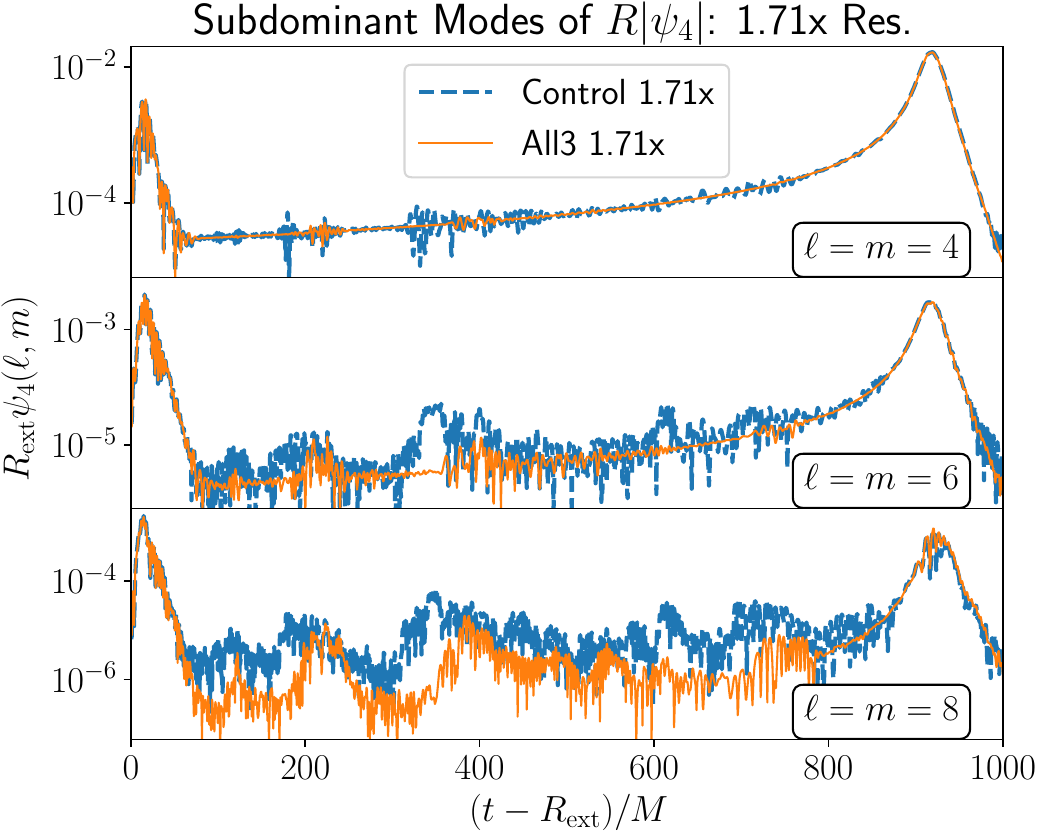}
  \end{center}
  \vspace{-0.4cm}
  \captionsetup{width=0.5\textwidth}
  \caption{\textbf{Subdominant modes}: Amplitude of spin-weight $s=-2$ subdominant even $\ell=m$ modes of $\psi_4$ up to 8, versus retarded time, at highest resolution. \textbf{Top panel}: $\ell=m=4$ mode, \textbf{Middle panel}: $\ell=m=6$ mode, \textbf{Bottom panel}: $\ell=m=8$ mode.}
  \label{fig:psi4noise_l6m6}
\end{figure}

Such significant noise reductions in the dominant $\psi_4$ mode are quite important, as GWs are usually the core payload of NR BBH simulations. Looking ahead, the improved sensitivities of current and future GW observatories will require significantly improved accuracy of NR GW predictions~\cite{Purrer:2019jcp,Ferguson:2020xnm}, including higher-order modes. In anticipation of this, Fig.~\ref{fig:psi4noise_l6m6} shows \allthree's impact on reducing noise in subdominant, even $\ell=m$ modes at the highest (1.71x) resolution. The noise at $\ell=m=4$ is greater in both \allthree and \control than at $\ell=m=2$, yet noise is still significantly reduced in \allthree. Turning to the $\ell=m=6$ subdominant mode, we find a truly exciting result. This mode is completely obscured by noise through the entire inspiral in \control, yet is mostly visible in \allthree, enabling the strain prediction to be extracted!

Still, there are limits to \allthree's effectiveness: the $\ell=m=8$ mode cannot be extracted from either \allthree or \control. This underscores the need for further research into improved techniques and numerical grid structures such as those that \Llama or \bhah provide.

\begin{figure}[ht!]
  \begin{center}
    \vspace{-0.cm}
    \includegraphics*[angle=0,width=0.48\textwidth]{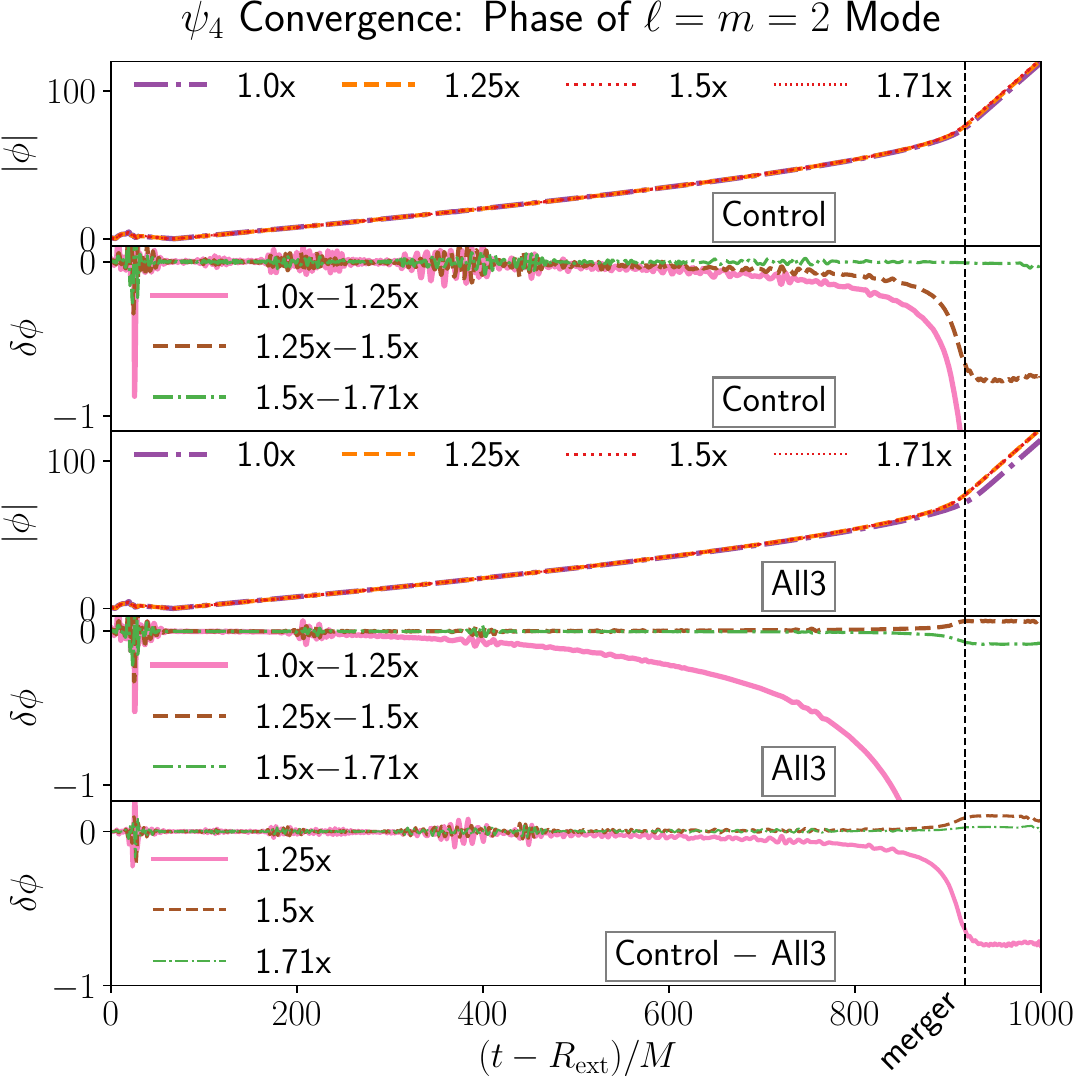}
  \end{center}
  \vspace{-0.4cm}
  \captionsetup{width=0.5\textwidth}
  \caption{\textbf{GW phase convergence analysis: $\ell=m=2$}. \textbf{Top panel}: accumulated phase $|\phi|$ of dominant ($\ell=m=2$) mode of $\psi_4$ at all four resolutions in \control. \textbf{Second panel from top}: Accumulated phase difference across neighboring resolutions in \control. \textbf{Third panel from top}: Same as top panel, but for \allthree. \textbf{Fourth panel from top}: Same as second panel, but for \allthree. \textbf{Bottom panel}: Accumulated phase difference between \control and \allthree at the three highest resolutions. All data were extracted from the simulation at extraction radius $R_{\rm ext} = 100M$.}
  \label{fig:psi4phaseconvergence}
\end{figure}

We next turn our attention to the convergence behavior of $\psi_4$, splitting our analysis into the phase and amplitude of the dominant $\ell=m=2$ mode. The phase in \control (top two panels of Fig.~\ref{fig:psi4phaseconvergence}) is too noisy to measure $n$ until a retarded time of about $780M$, when monotonic convergence behavior starts to emerge. However, although the implied convergence order of 2.3 between 1.0x, 1.25x, and 1.5x resolutions falls within the expected range of $2 \le n \le 8$, when comparing differences in 1.25x, 1.5x, and 1.71x resolutions, $n$ is roughly 12 before merger and 23 after merger, reflecting serious problems with convergence. We attribute this unusual behavior to the high-amplitude, stochastic oscillations in $\mathcal{H}$ between the BHs illustrated in Fig.~\ref{fig:binary_region_constraint_convergence}. As discussed in Sec.~\ref{subsec:BSSNconstraint}, these oscillations equate to spurious positive and negative effective energy densities, which, when they do not cancel, will impose an effective attraction or repulsion between the BHs, respectively. We fear that non-cancellations may directly impact the GW phase in potentially unpredictable ways.

Returning to Fig.~\ref{fig:binary_region_constraint_convergence}, $\mathcal{H}$ is roughly 100x lower between the BHs in \allthree as compared to \control. In spite of this, the persistent negative effective energy density at 1.0x resolution likely causes a significantly delayed merger in \allthree (middle panel, Fig.~\ref{fig:psi4phaseconvergence}). While this may seem troubling for \allthree, it is even more troubling for \control, as it implies that the stochastic behavior in $\mathcal{H}$ between the BHs would need to cancel to better than one part in 100 over the inspiral in order to exhibit similar or lower phase errors. Given the small number of waves observed in the binary region, such cancellation cannot be guaranteed, and is likely the cause of the inconsistent convergence in \control.

Considering \allthree at 1.25x and higher resolutions, Fig.~\ref{fig:binary_region_constraint_convergence} indicates that \allthree exhibits slight \textit{negative} oscillation about $\mathcal{H}=0$ in 1.25x, but a \textit{consistently positive} bias at 1.5x and 1.71x. Given inconsistent convergence observed in \control due to oscillations, as well as the clear negative bias of $\mathcal{H}$ in 1.0x \allthree results, caution is clearly warranted when considering \allthree 1.25x $\psi_4$ phase results. The fourth panel from the top of Fig.~\ref{fig:psi4phaseconvergence} indicates that while the difference between 1.25x and 1.5x phase results is smaller in \allthree than \control (second panel from the top), the difference flips sign close to merger, as compared to all other resolution pairs (\allthree or \control).

This likely implies that 1.25x resolution in \allthree and \control are slightly outside the convergent regime. While this analysis of $\mathcal{H}$ in the binary region (Fig.~\ref{fig:binary_region_constraint_convergence}) to explain non-convergent behavior has been useful, it remains to be seen whether it will be predictive of convergence in other BBH scenarios.

Next, we assess whether \control and \allthree waveforms converge to the same result. To this end, in the bottom panel of Fig.~\ref{fig:psi4phaseconvergence}, we fix the resolution and simply plot the phase differences between \control and \allthree. Notice that phase differences between \control and \allthree drop to tiny fractions of a radian with increasing resolution, indeed consistent with them converging to the same result.

\begin{figure}
  \begin{center}
    \vspace{-0.cm}
    \includegraphics*[angle=0,width=0.48\textwidth]{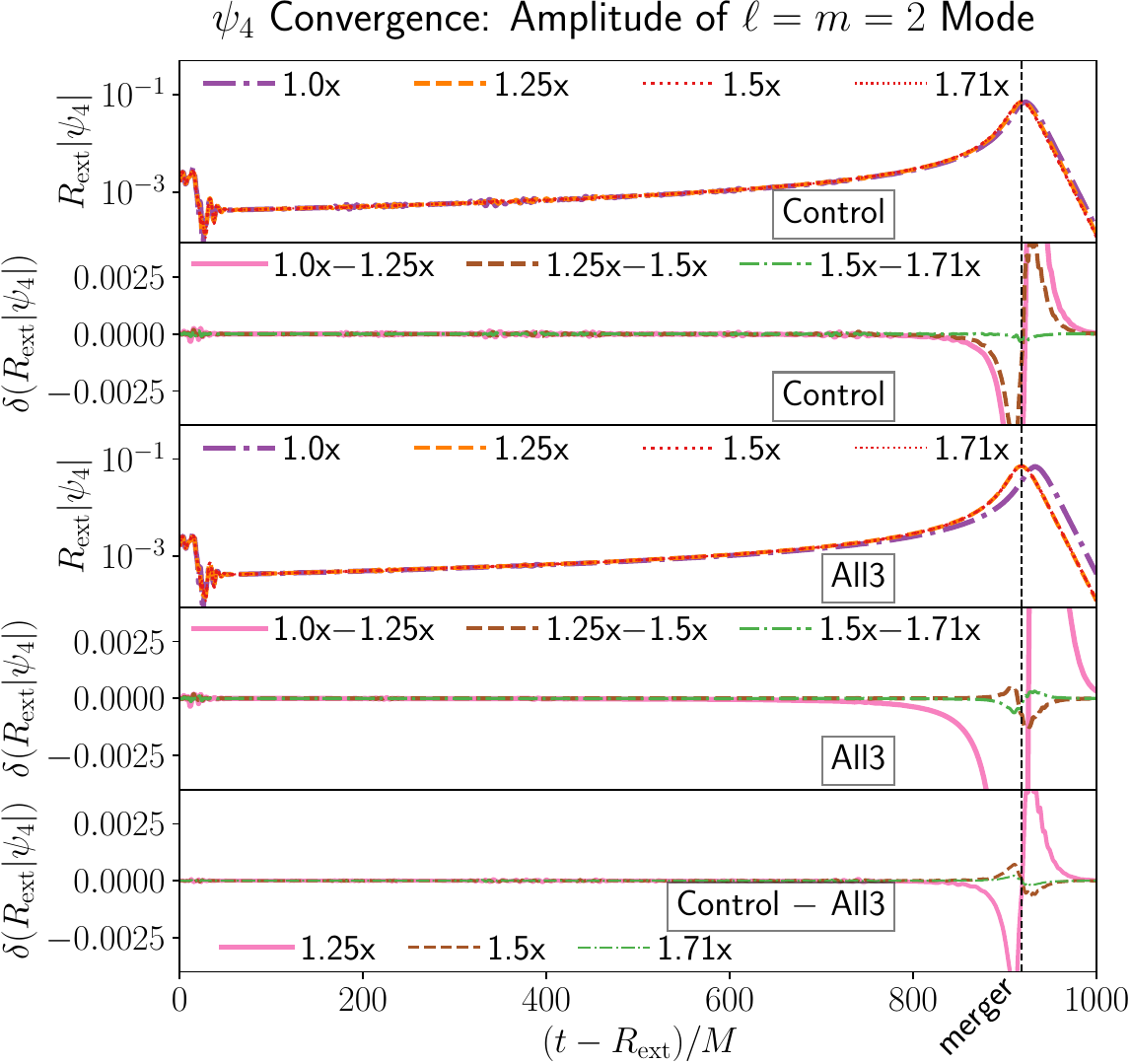}
  \end{center}
  \vspace{-0.4cm}
  \captionsetup{width=0.5\textwidth}
  \caption{\textbf{GW amplitude convergence analysis}. Same as Fig.~\ref{fig:psi4phaseconvergence}, except instead of phase, we consider the rescaled amplitude $R_{\mathrm{ext}}|\psi_4|$ of the dominant ($\ell=m=2$) mode of $\psi_4$.}
  \label{fig:psi4ampconvergence}
\end{figure}

For completeness, we present the same analysis for the rescaled amplitude of the extracted GW signal: $R_{\rm ext}|\psi_4|$, for the dominant mode in Fig.~\ref{fig:psi4ampconvergence}. Perhaps unsurprisingly, the conclusions drawn from our phase analysis hold for the amplitude as well, though the significantly delayed merger in the 1.0x resolution \allthree case is much more evident when considering the amplitude.

We conclude that improvements do not appear to negatively impact the dominant mode of the GW prediction; both amplitude and phase appear to converge to the same values as \control. That said, 1.25x resolution \allthree amplitude and phase results are much closer to the converged solution than in 1.25x \control, implying smaller errors at resolutions bordering the non-convergent regime. Perhaps most importantly, we find that \allthree's enormous reduction in Hamiltonian constraint noise in the binary region and the associated purely positive or negative biases, may enable us to predict for other BBH scenarios whether the actual merger time should be earlier or later than the given resolution. If so, Hamiltonian constraint information in the binary region, combined with \allthree, may be used to improve GW predictions from moving-puncture BBH merger simulations in the future.

\section{Conclusions and Future Work}
\label{sec:conclusion}

We have introduced three improvements to the moving-puncture technique, aimed at improving the accuracy of BBH evolutions. Our results show that the proposed improvements effectively reduce numerical errors and noise in different parts of the computational domain, from the strong-field area near the black holes to the weak-field region where GW predictions are extracted.

When combining all three improvements (\allthree), performance is significantly better than any single enhancement, cutting down Hamiltonian (momentum) constraint violations in the strong-field area by roughly two (three) orders of magnitude and in the GW-extraction zone by five (two) orders of magnitude. Further, the convergence of constraint violations improves near and inside the convergent regime, suggesting that these improvements become even more pronounced at higher resolutions.

The improvements also slightly reduce coordinate eccentricity and lower numerical noise in the dominant $\ell=m=2$ mode of $\psi_4$ by an average factor of 4.3 across different resolutions. Additionally, the fidelity of higher-order modes increases significantly, making the $\ell=m=6$ mode, which was previously hidden by noise, detectable. This is particularly encouraging given the demands of next-generation GW observatories.

Looking forward, we plan to explore the impacts of these improvements on BNS simulations and compare their efficacy to Z4-based formalisms. However, there are limitations to the wider applicability of these improvements: As the development of \cahd was built upon characteristic analysis of BSSN, it is unlikely to be useful in Z4-based formalisms, which exhibit a different characteristic structure and include damping terms already. Further, the initial gauge pulse is likely significantly less in BNS evolutions, meaning that \ssl may not be particularly helpful in this case.

Moving beyond Cartesian AMR-based grids, it will be interesting to learn the impacts of these improvements on mixed AMR/cubed-spheres grid structures like \Llama~\cite{LlamaCode}, as they may act to eliminate reflections and have already proven quite effective in \bhah. Meanwhile, work will continue to improve upon the moving-puncture technique further, with particular emphasis on addressing spikes in constraint violation associated with merger.

\section{Acknowledgments}
\label{sec:acks}

We would like to thank S.~Cupp, P.~Diener, R.~Haas, B.~J.~Kelly, and L.~R.~Werneck for their comments on early drafts of this manuscript. We would also like to thank J.~G.~Baker, W.~K.~Black, B.~Br\"ugmann, W.~East, E.~Hirschmann, C.~O.~Lousto, D.~Neilsen, V.~Paschalidis, and Y.~Zlochower for helpful discussions in the preparation of this manuscript. Z.B.E. gratefully acknowledges support from NSF awards PHY-2110352, OAC-2227105, AST-2227080, as well as NASA awards ISFM-80NSSC18K0538 and TCAN-80NSSC18K1488. This research made use of the resources of the High Performance Computing Center at Idaho National Laboratory, which is supported by the Office of Nuclear Energy of the U.S. Department of Energy and the Nuclear Science User Facilities under Contract No. DE-AC07-05ID14517. In addition, it made use of the Falcon~\cite{IdahoC3Plus3_2022_Falcon} supercomputer, operated by the Idaho C3+3 Collaboration.

\bibliographystyle{apsrev4-1}

\bibliography{references}

\end{document}